\documentclass[letter,11pt]{article}
\usepackage{hyperref}
\usepackage[affil-it]{authblk}
\usepackage{float}
\usepackage{xcolor}
\usepackage{graphicx}
\usepackage[title]{appendix}
\usepackage{amssymb}
\usepackage{amsthm}
\usepackage[margin=1in]{geometry}

\title{MAISE: Construction of neural network interatomic models and evolutionary structure optimization}

\author[a]{Samad Hajinazar}
\author[a]{Aidan Thorn}
\author[a]{Ernesto D. Sandoval}
\author[a]{Saba Kharabadze}
\author[a,*]{Aleksey N. Kolmogorov}
\affil[a]{Department of Physics,
  Applied Physics and Astronomy, Binghamton University, State
  University of New York, PO Box 6000, Binghamton, New York
  13902-6000, United States}
\affil[*]{Corresponding author: kolmogorov@binghamton.edu}

\date{}                     
\begin{document}

\maketitle

\begin{abstract}

 Module for \textit{ab initio} structure evolution (MAISE)
 is an open-source package for materials modeling and
prediction. The code's main feature is an automated generation of
neural network (NN) interatomic potentials for use in global structure
searches. The systematic construction of Behler-Parrinello-type NN
models approximating \textit{ab initio} energy and forces relies on
two approaches introduced in our recent studies. An evolutionary
sampling scheme for generating reference structures improves the NNs'
mapping of regions visited in unconstrained searches, while a
stratified training approach enables the creation of standardized NN
models for multiple elements. A more flexible NN architecture proposed
here expands the applicability of the stratified scheme for an
arbitrary number of elements. The full workflow in the NN development
is managed with a customizable 'MAISE-NET' wrapper written in
Python. The global structure optimization capability in MAISE is based
on an evolutionary algorithm applicable for nanoparticles, films, and
bulk crystals. A multitribe extension of the algorithm allows for an
efficient simultaneous optimization of nanoparticles in a given size
range. Implemented structure analysis functions include fingerprinting
with radial distribution functions and finding space groups with the
SPGLIB tool. This work overviews MAISE's available features,
constructed models, and confirmed predictions.

\end{abstract}


\section{Introduction}
\label{Sintro}


Development of advanced modeling methods and simulation tools
continues to reshape the process of materials discovery and
characterization
\cite{CurtaroloNature,MatPredreview2019,NNreview2019Csanyi,Oganov2019}. In
structure prediction, unconstrained optimization algorithms enable
identification of thermodynamically stable phases with no prior
information and have been widely used to guide the experimental work
\cite{Oganov2019,bh2,Woodley2008}. The challenge of finding
global minima in large configuration spaces has been addressed with a
variety of strategies: basin-hopping \cite{bh0} represents an
efficient procedure for escaping from local minima and mapping the
potential energy surface (PES), particle swarm optimization relies on
the crowd intelligence for navigating the energy landscape
\cite{Eberhart1995}, random searching provides an unbiased
configuration sampling \cite{AIRSS}, evolutionary algorithm mixes and
propagates beneficial structural traits
\cite{ganp0,ganp1,softmutation}, etc. In interatomic interaction
modeling, machine learning frameworks have emerged as attractive
alternatives to traditional potentials
\cite{PhysRevLett.98.146401,PhysRevLett.104.136403}. Numerous recent
studies have been dedicated to improving the methodology for
representing atomic environments, generating reference data sets, and
training machine learning models.
\cite{PhysRevLett.98.146401,PhysRevLett.104.136403,Si-HP,PhysRevB.81.184107,NNcharge,PhysRevB.83.153101,PhysRevB.85.045439,PSSB:PSSB201248370,Zmorawietz2013full,0953-8984-26-18-183001,C4CP04751F,doi:10.1063/1.4966192,AENET,Khorshidi2016310,AK34,force1,grandreview17,csanyi-C,PROPhet,SchNet,AK37,Pt13,csanyi-NNsemi,csanyi-B,DeepPot-SE,AK38,AK40,NNreview2019,NNreview2019Csanyi,SIMPLE-NN,PANNAcode,FLAMEcode,NNreviewMODELS2019,NN-Si,POETcode,Lomaka,tensoralloy}.

The aim of this study is to introduce the main materials modeling
capabilities available in our module for \textit{ab initio} structure
evolution (MAISE), starting with a guide into the package installation
(Section \ref{Scompile}) and basic features (Sections \ref{Sflags} and
\ref{Ssim}). Given a number of excellent reviews detailing the
background on unconstrained structure prediction
\cite{Woodley2008,Oganov2019} and machine learning
\cite{doi:10.1063/1.4966192,NNreview2019Csanyi}, our presentation
focuses on describing MAISE's distinctive algorithms, functionalities,
and applications previewed in the following paragraphs.

MAISE was first written as a standalone C code in 2009 \cite{AK16}. It
was originally designed as an evolutionary optimization engine
interfaced with external density functional theory (DFT) packages to
enable unconstrained ground state structure searches. The implemented
evolutionary algorithm  (Section \ref{Sevo})  followed a general
principle of using natural selection to evolve populations of
structures with crossover and mutation operations
\cite{ES-300,ES-200,USPEXcode,Oganov,ES-999,ganp0,ganp3,ES-600,EVOcode,GASPcode,zunger,oganov1,MUSEcode,oganov2,XTALOPTcode}. MAISE-specific
features include radial distribution function (RDF)-based structure
fingerprinting for detecting and eliminating similar population
members \cite{AK16,AK23,AK25} and an efficient co-evolutionary
optimization of nanoparticles (NP) in a given size range via sharing
of best motifs among multiple tribes \cite{AK38,AK40}. \textit{Ab
  initio} predictions made with MAISE and confirmed in experimental
studies are overviewed in Section \ref{Sstr}.

The primary function of the present MAISE package is the construction
of neural network  (NN)  interatomic models for accurate mapping of
\textit{ab initio} PES's. Our examinations of NN performance in
prediction of stable compounds have revealed limitations of the
traditional approaches used to sample configuration spaces and train
NNs for multiple elements \cite{AK34}. An evolutionary sampling and a
stratified training schemes introduced in Ref. \cite{AK34} and
discussed in Section \ref{Snn} have allowed us to build reliable NN
models for extended sets of metals. Our developed MAISE-NET Python
script streamlines all stages of the process, from generating
reference structures and handling external \textit{ab initio}
calculations to performing NN training and testing  (see Section
\ref{Smaisenet}). The library of the latest generation of NN models
constructed with the MAISE-NET script are described in Section
\ref{Slib}. The efficiency of NN calculations, the performance of NN
models, and the first NN-based structure predictions are described in
Section \ref{Snntest}. With the machine learning module and relevant
utility functions comprising about 9,130 out of 14,364 lines of the
full code, a more descriptive reading of the MAISE acronym at this
point is 'module for artificial intelligence and structure evolution'.

 MAISE command-line structure analysis and manipulation operations,
such as structure comparison or space group determination, are listed
in Section \ref{Sflags}. The code can perform local/global
optimizations, molecular dynamics (MD), and basic phonon calculations
by evaluating the total energy, atomic forces, and unit cell stresses
for given structures at the NN or empirical potential levels (see
Section \ref{Ssim}). The main input/output files have a general VASP
\cite{VASP1,VASP2} format to simplify interfacing MAISE with other
structure prediction and property analysis engines (PyChemia
\cite{FireflyPyChemia}, PHON \cite{PHON}, etc.). The NN training and
structure simulation modules are parallelized with OpenMP
\cite{OpenMP}.

\section{Installation}
\label{Scompile}

{\it Download} The full MAISE package, currently MAISE version 2.5 and
MAISE-NET version 1.0, can be obtained from the Github repository
\cite{MAISE,MAISE-NET}. It contains MAISE C-language source code,
MAISE-NET Python script (Section \ref{Smaisenet}), available NN and
empirical potential models (Section \ref{Slib}), and basic examples.

{\it Compilation} The source code for MAISE can be compiled with: {\$
  make -\phantom{}-jobs}. During MAISE compilation, the makefile
script checks if two required external libraries, GSL \cite{GSLRef}
and SPGLIB (v1.11.2.1, Feb 2019) \cite{SPGLIBRef}, are present. If
not, they will be automatically downloaded to ./ext-dep and installed
in ./lib on most systems. If the GSL or SPGLIB installation is not
completed automatically the user should compile them manually and copy
(i) libgsl.a, libgslcblas.a and libsymspg.a into the './lib'
subdirectory; (ii ) the 'spglib.h' header into './lib/include'
subdirectory; and (iii) all gsl headers into the './lib/include/gsl'
subdirectory.

{\it Post-compilation test} A 'check' script is available in the
'./examples/' directory which can be run after compiling the MAISE
executable to ensure the proper function of the code. The script
parses a small dataset, trains a basic NN, and optimizes a crystal
structure. Error logs are generated in case any issues are detected.

\section{Unit cell analysis and manipulation}
\label{Sflags}

A variety of structure analysis and manipulation tools are implemented
in MAISE package which can be used in the command-line with the
corresponding task-specifier flag. Working primarily with the VASP
structure format (POSCAR file) as input, MAISE can determine the space
group, calculate the radial distribution function (RDF)
\cite{AK16,AK23}, measure the similarity of two structures via RDF
pattern comparison, calculate volume per atom for bulk and cluster
geometries \cite{npann1}, align the cluster in the simulation box
along the high symmetry axes, etc. The code expects a 'POSCAR' file in
the running directory for operations involving a single structure or
two 'POSCAR0' and 'POSCAR1' files for structure comparison. The tasks
listed in Table \ref{Tflags} can be performed in the command line by
running: {\$ maise -[flag]}.

The similarity, or dot product, between two structures $k=1,2$ with
species $N_{\mathrm{spc}}$ has been defined in MAISE as

\begin{eqnarray}
C_1\cdot C_2 =
\sum_n^{N_{\mathrm{bin}}}\sum_{s1}^{N_{\mathrm{spc}}}\sum_{s2}^{N_{\mathrm{spc}}} \mathrm{RDF}_{1,s1,s2}(R_n)\mathrm{RDF}_{2,s1,s2}(R_n)/(\mathrm{norm}_1\mathrm{norm}_2),
\nonumber
\end{eqnarray}
\begin{eqnarray}
\mathrm{norm}_k = \left[ \sum_n^{N_{\mathrm{bin}}}\sum_{s1}^{N_{\mathrm{spc}}}\sum_{s2}^{N_{\mathrm{spc}}} \mathrm{\
RDF}_{1,s1,s2}(R_n)\mathrm{RDF}_{2,s1,s2}(R_n)  \right]^{1/2}.
\nonumber
\end{eqnarray}
The RDFs are defined for each structure $k$ at each bin $R_n=n/N_{\mathrm{bin}}R_{\mathrm{hard}}$ ($N_{\mathrm{bin}}=3,000$) as
\begin{eqnarray}
\mathrm{RDF}_{k,s1,s2}(R_n)=\sum_{i,si=s1}^{N_{\mathrm{atom}}}\sum_{j\neq
  i,sj=s2}^{N_{\mathrm{atom}}}e^{-\frac{(R_{ij}-R_n)^2}{2\sigma^2}}
  f_{\mathrm{cut}}(R_n), \nonumber 
\end{eqnarray}
where $si$ and $sj$ denote the species of atoms $i$ and $j$,
respectively. $f_{\mathrm{cut}}(R_n)=1$ for $R_n<R_{\mathrm{soft}}$
and
$f_{\mathrm{cut}}(R_n)=\cos\left(\pi/2\frac{R_n-R_{\mathrm{soft}}}{R_{\mathrm{hard}}-R_{\mathrm{soft}}}\right)$
for $R_{\mathrm{soft}}<R_n<R_{\mathrm{hard}}$. For efficiency purposes,
only $R_n-3\sigma<R_{ij}<R_n+3\sigma$ are included in the sum.

The dot product is sensitive to the choice of $R_{\mathrm{soft}}$,
$R_{\mathrm{hard}}$, and $\sigma$. It is good practice to include at
least two shells of nearest neighbors ($R_{\mathrm{hard}}\gtrsim 5$
\AA) and use sharper Gaussians ($\sigma\approx0.008$ \AA) for
disordered or cluster structures and wider ones for high-symmetry
structures ($\sigma\approx0.02$ \AA).

\begin{table}[h]
\centering
\begin{tabular}{p{0.1\columnwidth}|p{0.8\columnwidth}}\hline\hline
flag &  description            \\ \hline
man  &  output the list of available flags \\ 
rdf  &  compute the RDF for POSCAR \\ 
cxc  &  compute dot product for POSCAR0 and POSCAR1 using RDF \\ 
cmp  &  compare RDF, space group, and volume of POSCAR0 and POSCAR1 \\ 
spg  &  convert POSCAR into str.cif, CONV, PRIM \\ 
cif  &  convert str.cif into CONV \\ 
rot  &  rotate a cluster along eigenvectors of moments of inertia \\ 
dim  &  find whether POSCAR is periodic (3) or non-periodic (0) \\ 
box  &  reset the box size for clusters \\ 
sup  &  make a $N_a\times N_b \times N_c$ supercell\\ 
vol  &  compute volume per atom for crystals or clusters \\ \hline\hline
\end{tabular}
\caption{List of the available command-line flags in MAISE package for
  structure analysis and manipulation.}
\label{Tflags}
\end{table}

\section{Structure simulation}
\label{Ssim}

Available structure simulation functions include unit cell
relaxation, MD, and phonon property analysis. The
structure, the interaction model, and the job settings are specified
in 'POSCAR', 'model', and 'setup' files, respectively.

\subsection{Local structure optimization}

Structure optimization with analytic derivative-based BFGS
\cite{BFGS} or CG \cite{CG} algorithms can be performed by using NN or
other classical interatomic interaction models available in MAISE. The
local optimization is carried out until the maximum number of
iterations (MITR) or the targeted enthalpy difference between
successive steps (ETOL) is reached. The full list of relevant 'setup'
parameters for the local optimization task is provided in Table
\ref{Trelaxsetup}.

The unit cell parameters, total/atomic energies, and force/stress
components can be outputted at each relaxation step in an 'OUTCAR'
file, while the final structure is saved in a 'CONTCAR' file. This
information saved in the VASP-style format can be utilized by external
codes to perform vibrational property analysis, global structure
optimization, etc.

\subsection{Molecular dynamics simulations}

MD simulations can be run in the microcanonical ensemble ($NVE$) with
the Verlet algorithm \cite{Verlet}, the canonical ensemble ($NVT$)
with the Nos\'{e}-Hoover thermostat \cite{Nose,Hoover}, and
isobaric-isothermal ensemble ($NPT$) with a combination of the
Nos\'{e}-Hoover thermostat and the Berendsen barostat
\cite{Berendsen}. The velocities are initialized either according to
the Maxwell distribution at a given starting temperature or with the
values specified in the 'POSCAR' file. Table \ref{Tmd} lists 'setup'
parameters relevant for MD simulations. MAISE outputs energies,
lattice parameters, Lindemann index, average RDF, etc. for each
temperature. In the current version of MAISE, Lindemann index value is
well-defined only for NPs and the barostat is implemented for unit
cells with orthogonal lattice vectors.

\begin{figure}[h]
\centering
\includegraphics[width=0.8\textwidth]{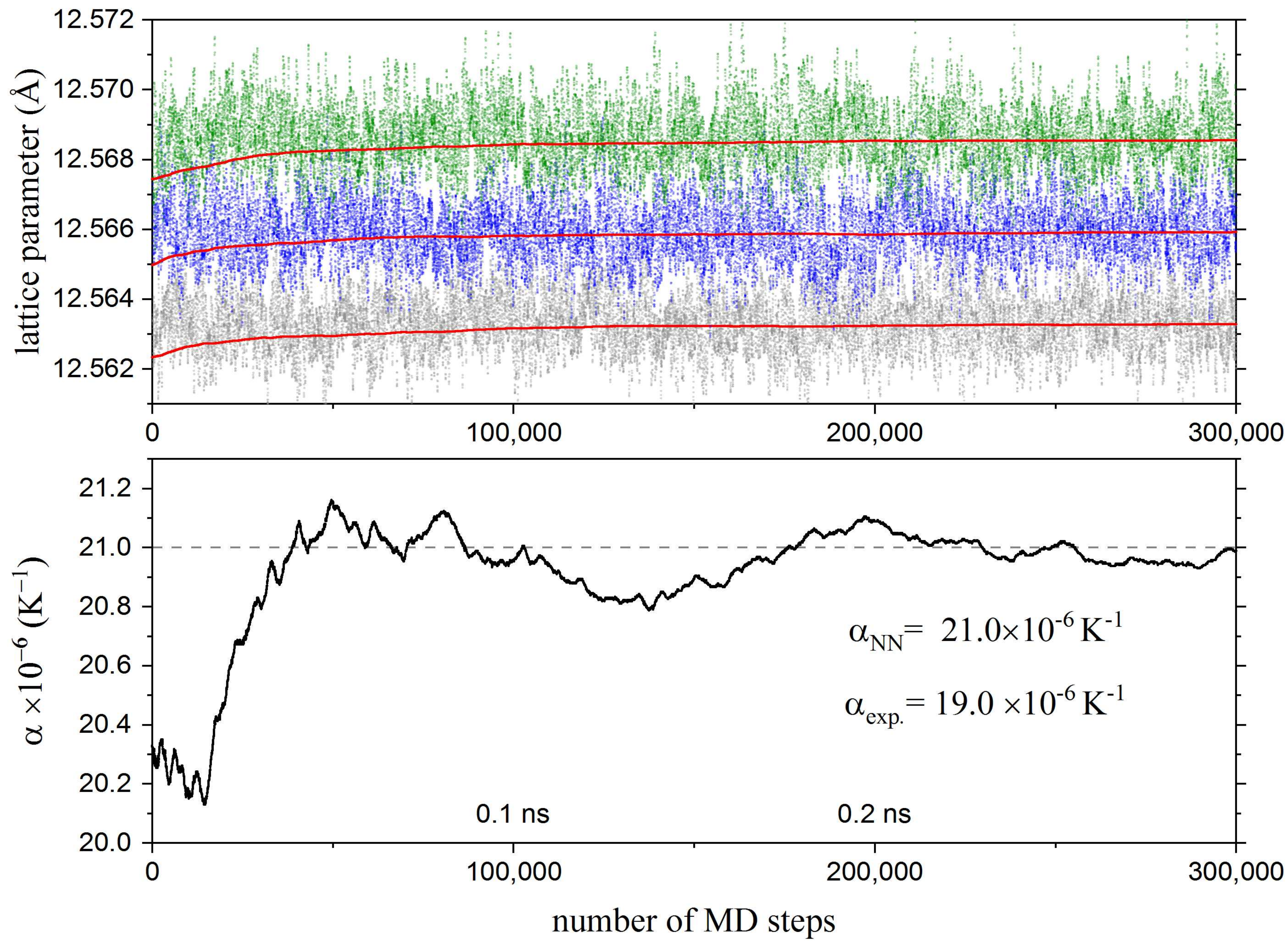}
\caption{(Top panel) Fluctuations of the lattice parameter for a
  108-atom supercell of FCC-Ag at $T=290$ K (grey), $T=300$ K (blue),
  and $T=310$ K (green) along with the corresponding average lattice
  parameters (red) as a function of the number of MD steps. (Bottom
  panel) Linear thermal expansion coefficient ($\alpha$) at $T=300$ K
  as a function of the number of MD steps. The calculated linear
  expansion coefficient for Ag is in 10\% agreement with the measured
  value \cite{AIPHandbook}.}
\label{Pmd}
\end{figure}

Figure \ref{Pmd} illustrates the use of the $NPT$ ensemble and our
latest NN model for evaluating the linear thermal expansion
coefficient $\alpha=\frac{1}{L}\left(\frac{\partial L}{\partial
  T}\right)_P$ in Ag near room temperature. A 108-atom
$3\times3\times3$ supercell of FCC-Ag was simulated at $T=300\pm 10$ K
for 0.5 ns with a 1 fs time step (500,000 integration steps in total)
to find the numerical temperature derivative of the lattice
constants. Allowing the first 0.025 ns for equilibration, we observed
convergence of $\alpha$ to within 0.5\% in the following 0.25
ns. Simulations of FCC-Cu and BCC-Na showed similar convergence
rates. The resulting linear thermal expansion coefficients of
$21.0\times10^{-6}$ K$^{-1}$ for Ag, $14.9\times10^{-6}$ K$^{-1}$ for
Cu, and $51.7\times10^{-6}$ K$^{-1}$ for Na are within 10-30\%
relative to the corresponding experimental values of
$19.0\times10^{-6}$ K$^{-1}$, $16.7\times10^{-6}$ K$^{-1}$, and
$69\times10^{-6}$ K$^{-1}$ \cite{AIPHandbook}. Simulations with a
smaller temperature difference $T=300\pm 5$ K and a larger structure
(256-atom $4\times4\times4$ supercell of FCC-Ag) showed similar
results for the expansion coefficient.

\subsection{Phonon calculations}

Our studies of vibrational properties \cite{AK40} have been performed
with an external PHON package \cite{PHON} because it readily links
with VASP or MAISE for a consistent comparison of the NN models
against the DFT. Presently, MAISE has an internal option to calculate
$\Gamma$-point phonons with the frozen phonon method in the
quasi-harmonic approximation. The dynamical matrix is constructed by
numerical differentiation of the atomic forces. The magnitude of
atomic displacements of each atom is defined by the 'DISP'
parameter. Due to the negligible numerical noise of the NN analytic
forces, the displacement values can be kept small to reduce the
anharmonic effects and satisfy the acoustic sum rule (a list of setup
parameters for phonon calculations in MAISE code is presented in Table
\ref{Tphonsetup}).

The main application of this basic feature is to determine the
presence of soft frequencies in the analysis of structures' dynamical
stability. The code marks trivial zero-frequency translational (and
rotational) modes by checking whether the eigenvectors generate net
linear (and angular) momenta in crystals (and clusters). Ordered
frequencies and the corresponding eigenvectors are printed in the
'OUTCAR' file and can be used for introducing soft-mode mutations in
global evolutionary searches \cite{softmutation} or monitoring nudged elastic band
method convergence in transition state searches.

\section{Evolutionary search}
\label{Sevo}

{\it Overview} Evolutionary algorithms rely on Nature's heredity and
'survival of the fittest' principles for optimizing complex
systems. MAISE enables the search for lowest-enthalpy bulk crystals,
flat films, or NPs at a fixed chemical composition. The
majority of the algorithm's numerous internal parameters related to
the generation, evolution, and selection of structures have been tuned
for typical crystalline unit cells with up to about 50 atoms and
NPs with a few hundred atoms. Below we briefly overview the
key settings adjustable by the user for the algorithm's optimal
performance.

{\it Interaction description method} The evolutionary
optimization module expects local relaxations of structures to be
performed by an external code (flag CODE) through a queueing system
(flag QUET). The current version is linked with VASP for DFT
calculations and with MAISE for NN calculations. In case of fast
Lennard-Jones, Gupta, or Sutton-Chen potentials, local optimization
calls can be made directly from the evolutionary engine in
MAISE. Input files and submission scripts for DFT/NN relaxations
should be specified in the INI directory.

{\it Population initialization} Bulk ground state searches can be
initialized via (i) randomization of given structures to bias the
search toward nearby stable configurations; (ii) randomization of
atoms in a constrained unit cell to make use of available information
from XRD; and (iii) unbiased generation of random unit cells and
atomic positions. In case the structures have interatomic distances
shorter than a tabulated species- and pressure-dependent value,
they are adjusted using a simple repulsive interatomic potential or
re-generated. NPs can be created with a TETRIS-like function
introduced in our recent study \cite{AK38} that ensures good packing
and customizable radial/angular distributions of species. 2D films are
constrained to the $x$-$y$ plane at the beginning and duration of the
ES \cite{AK33}.

{\it Evolution operations} Offspring bulk structures are obtained with
mutation or crossover operations. The former acts on a randomly chosen
parent structure to distort lattice vectors, displace atomic
positions, and/or swap atoms of different species. The latter randomly
picks two parent structures, rotates the lattice vectors to ensure the
best matching of unit cell dimensions, slices the unit cells
approximately in half, and combines the pieces with small adjustments
at the boundary to avoid short interatomic distances. Offspring
NPs can also be created with alternative ``Rubik's Cube''
and ``spherical cut crossover'' operations, described in our previous
study and used to quantify the effectiveness of the traditional
crossover \cite{AK38}.

{\it Structure selection} Once a new generation is locally optimized,
the joint population of parent and child structures is ranked
according to their enthalpy and each structure $n$ is assigned the
survival probability proportional to $1/2
(1-\tanh[2(H_{n}-H_{min})/(H_{max}-H_{min})-1])$ where $H_{min}$ and
$H_{max}$ are lowest and highest enthalpies in the population,
respectively \cite{ES-901}. Duplicate structures determined to have
similar RDFs, energies, and volumes are assigned zero chance of
survival. Structures are eliminated one by one until the merged
population is reduced to its original size. Ground states with 10-16
atoms per primitive unit cells are usually found in 1,000-3,000 local
optimizations. Configurations with large lattice constant differences
(e.g., long stacking sequences) and low atomic densities (e.g., the
low-coordination diamond structure) tend to take longer to appear.

{\it Job execution} The evolutionary optimization is executed by
running MAISE in the background. The search (re)starts from a given
generation and proceeds for a specified number of iterations
(flag NITR). In each cycle, the code generates a new population, submits a
job for each structure to a specified queue, checks if the jobs
finished successfully, processes the results, and outputs
enthalpy/volume for each structure.

{\it Multitribe optimization} The efficient co-evolutionary
simultaneous optimization of NPs introduced and tested in
our recent study \cite{AK40} requires a separate bash script. The
wrapper manages the submission of ESs and the exchange of seeds among
tribes at the end of each cycle of isolated evolution.

\begin{figure}[h]
\centering
\includegraphics[width=\textwidth]{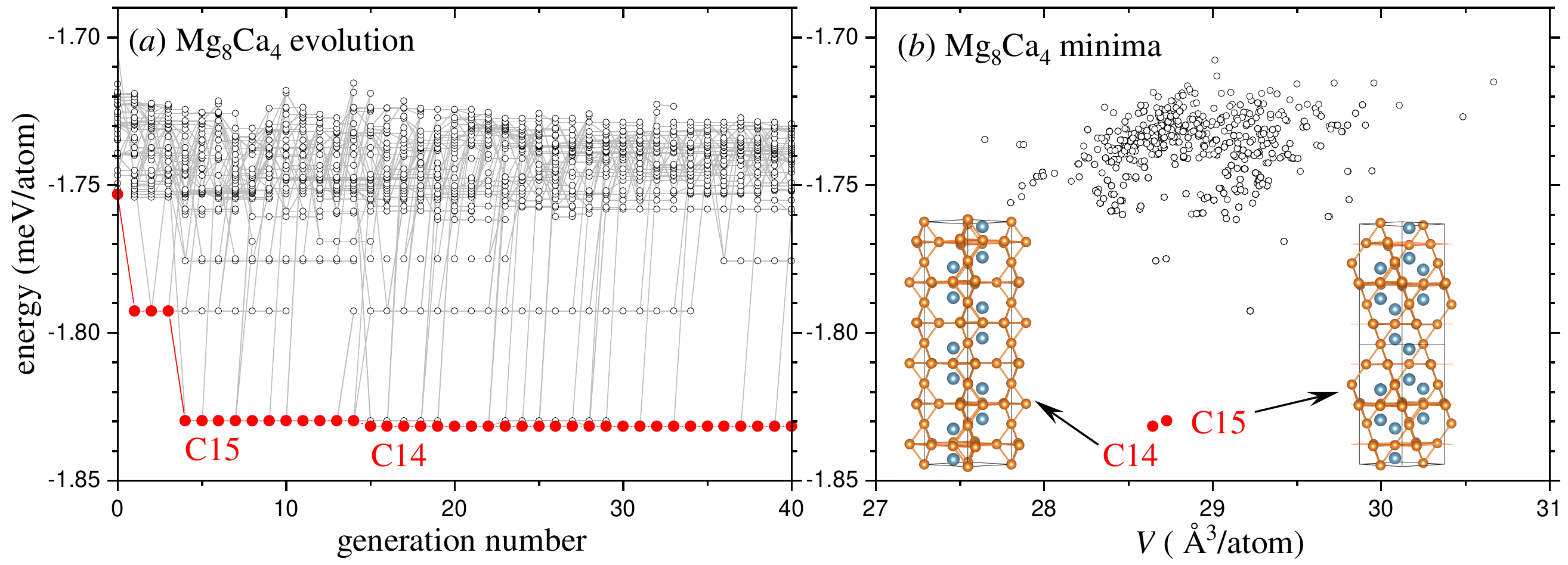}
\caption{Typical results of ES runs performed with MAISE. This global
  structure search at the Mg$_8$Ca$_4$ composition identified both the
  metastable C15 Laves phase and the C14 ground state. The
  interactions were modeled with the latest Mg-Ca NN interatomic
  potential. (a) Energy distribution and structure heredity for an ES
  with 40 generations and 32 members in the population. (b) Collection
  of all local/global minima at the end of the ES.}
\label{Fevos}
\end{figure}

{\it ES output} The ES progress can be monitored by visualizing the
enthalpy profile and heredity of population members saved in
'ebest.dat', 'erank.dat', and 'elink.dat' files (see
Figure \ref{Fevos}). The connections between points in consequent
generations illustrate which parent structures were used to generate
the offspring: one for mutations and two for crossovers. After an ES
is completed, one can select distinct low-enthalpy structures in the
entire pool of locally optimized members by running a post-search
analysis with JOBT=13. Configurations with dissimilar RDF dot products
by at most SCUT (e.g., 0.95) and with enthalpies at most DENE (e.g.,
20 meV/atom) above the lowest-enthalpy structure will be saved and
optionally relaxed at the DFT level.

{\it ES example} Figure \ref{Fevos} illustrates the performance of a
typical ES. Structures with Mg$_8$Ca$_4$ unit cells were modeled with
our latest NN interatomic potential. A population of 32 members was
evolved for 40 generations and converged to the known C14 Laves phase
ground state, producing the metastable C15 along the way.

\section{Confirmed  {\it ab initio}  predictions}
\label{Sstr}

The reliability of \textit{ab initio} predictions for finding new
materials depends on the accuracy of the theoretical method for
computing the structure stability (Gibbs free energy) and the
exhaustive sampling of large configuration spaces (structures and
compositions). A common approach to evaluating Gibbs free energy with
continually improving DFT approximations
\cite{LDA,PBE,LDA+U1,LDA+U2,vdW1, PhysRevLett.91.126402,SCAN} is to
determine the enthalpy at $T=0$ K and then include the
temperature-dependent vibrational/configurational entropy terms for
viable candidates. Explorations of configurational spaces can be done
with a variety of advanced structure prediction methods introduced in
the past two decades
\cite{ganp0,ganp3,ganp1,ganp2,dls0,dls1,dls2,MH,AIRSS,bh0,bh1,bh2,ps0,ps1,Wu2017}. The
search strategy employed in our predictive work has involved (i)
high-throughput (HT) screening of known relevant prototypes to
establish a baseline for compound stability; (ii) unconstrained
evolutionary search (ES) to identify new stable motifs; and (iii)
stability analysis to explain or improve the stability of identified
materials.

\begin{figure}[h]
\centering
\includegraphics[width=0.8\textwidth]{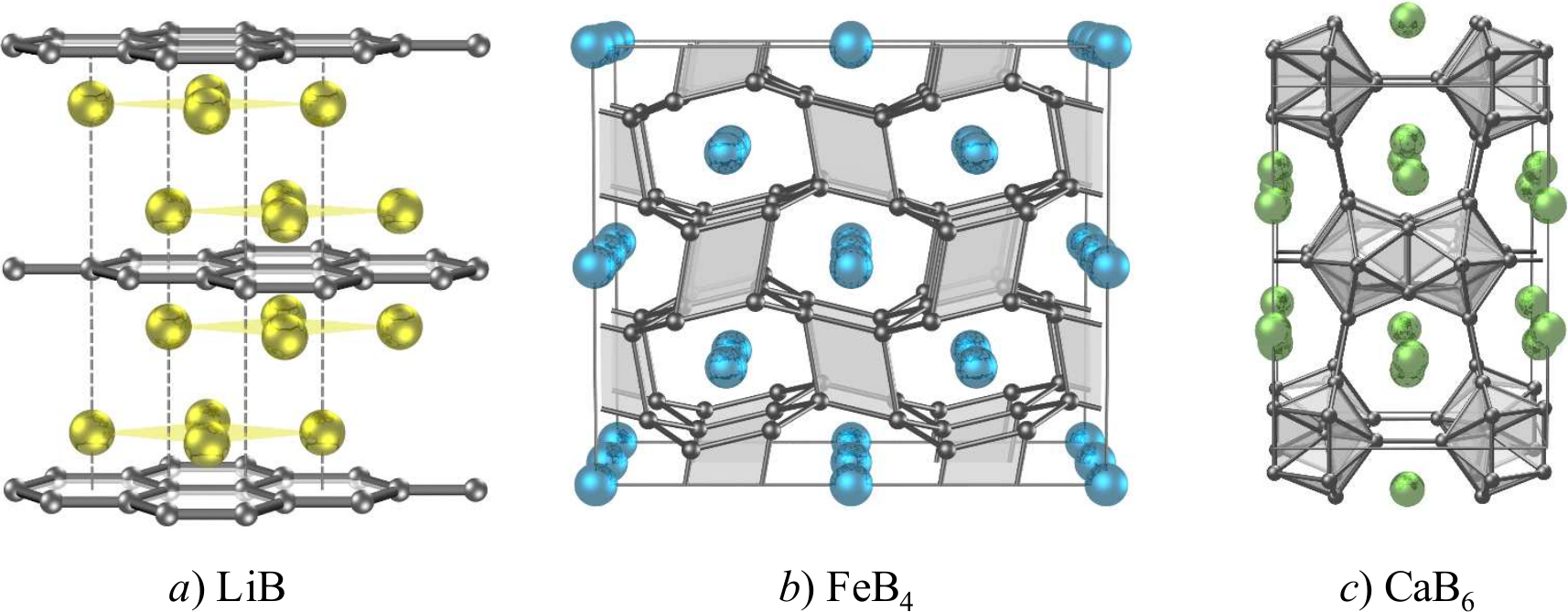}
\caption{Structures of select MAISE confirmed predictions detailed in
  Table \ref{Tstr}. The small (large) spheres show boron (metal)
  atoms.}
\label{Pstr}
\end{figure}

Here, we recount notable factors leading to successful predictions and
provide context on the discovered materials' significance (Figure
\ref{Pstr} and Table \ref{Tstr}). In terms of novelty, (i) FeB$_4$
\cite{AK16,AK17,PhysRevLett.111.157002}, LiB \cite{AK08,AK09,AK30},
and NaSn$_2$ \cite{AK31,doi:10.1021/jacs.7b01398} are new phases
predicted before their experimental realization; (ii) CaB$_6$
\cite{AK23} and Na$_3$Ir$_3$O$_8$ \cite{AK36} are solutions of complex
phases synthesized and characterized in joint studies; and (iii)
Na$_2$IrO$_3$ \cite{AK21}, CrB$_4$ \cite{AK17,AK22}, and MnB$_4$
\cite{AK28,Barbara2014,C4CP01339E} are confirmed revisions of
previously misidentified crystal structures. All cases except for
Na$_2$IrO$_3$ involved extensive evolutionary searches and resulted in
brand-new crystal structures for FeB$_4$, CrB$_4$, MnB$_4$, CaB$_6$,
and Na$_3$Ir$_3$O$_8$. All phases except for Na$_3$Ir$_3$O$_8$ have
been either synthesized at or successfully quenched down to the
ambient pressure.

FeB$_4$ \cite{AK16,AK17,PhysRevLett.111.157002} is an early example of
a superconductor predicted fully '{\it in silico}'. With a combination of HT
screening, ESs, and electron-phonon calculations, we demonstrated that
an FeB$_4$ compound should become thermodynamically stable under
moderate pressures around 10 GPa in a brand-new oP10 crystal structure
(SG\#58), remain metastable under normal conditions, and exhibit
phonon-mediated superconductivity unusual for an Fe-containing
material. The subsequent discovery of the superconductor
\cite{PhysRevLett.111.157002} has motivated further
studies \cite{FeB4_2,FeB4_1}.

LiB was proposed to be a new synthesizable layered phase
\cite{AK08,AK09} with electronic features desirable for MgB$_2$-type
superconductivity \cite{MgB2_exp}. The set of 'metal sandwich'
configurations was constructed by analyzing stability trends in our HT
data. In order to determine suitable synthesis conditions, we
explained the off-stoichiometric LiB$_x$ material and modeled the
complex behavior of the two competing phases under high pressures. Our
synthesis and XRD analysis confirmed the predicted shifts in the
LiB$_x$ composition and the existence of the LiB phase with random
stacking \cite{AK30}. The demonstration of the LiB metastability under
ambient pressure should simplify future study of the material's
superconductivity.

NaSn$_2$ \cite{AK31} was predicted to be an overlooked phase
synthesizable under ambient pressure. With the primary focus on
finding new bulk Sn materials that could be exfoliated into stanene,
we examined a set of layered Sn alloys and showed that Na stabilizes a
rigid 3D framework with flat Sn layers. Our electronic structure
analysis indicated that the compound should have non-trivial
topological properties. The predicted hP3-NaSn$_2$ phase (SG\#191) was
observed later in an independent experiment
\cite{doi:10.1021/jacs.7b01398}.

\begin{table}[t]
\centering
\begin{tabular}{p{0.15\columnwidth}p{0.15\columnwidth}p{0.15\columnwidth}|p{0.45\columnwidth}}\hline \hline
phase             & prediction                            & synthesis P & properties                            \\
structure         & confirmation                          & quenched P  & remarks                               \\ \hline
FeB$_4$           & 2010 \cite{AK16,AK17}                 & 10 GPa      & Fe-based BCS superconductor           \\
oP10              & 2013 \cite{PhysRevLett.111.157002}    & 1 bar       & predicted fully 'from scratch'        \\
                  &                                       &             &                                       \\
LiB               & 2006 \cite{AK08,AK09}                 & 21 GPa      & proposed MgB$_2$-type superconductor  \\
hP4-8             & 2015 \cite{AK30}                      & 1 bar       & cold compression synthesis            \\
                  &                                       &             &                                       \\
NaSn$_2$          & 2016 \cite{AK31}                      & 1 bar       & 3D Sn framework with flat Sn layers   \\
hP3 (AlB$_2$)     & 2017 \cite{doi:10.1021/jacs.7b01398}  &             & non-trivial topological properties    \\
                  &                                       &             &                                       \\
CaB$_6$           & 2012 \cite{AK23}                      & 31 GPa      & unique boron building blocks          \\
tI56              & 2012 \cite{AK23}                      & 1 bar       & found w/o any structural input        \\
                  &                                       &             &                                       \\
Na$_3$Ir$_3$O$_8$ & 2018 \cite{AK36}                      & 11 GPa      & dimerized Ir framework                \\
mP56              & 2018 \cite{AK36}                      & 1 bar       & found w/o any high-P structural input \\
                  &                                       &             &                                       \\
Na$_2$IrO$_3$     & 2012 \cite{AK21}                      & 1 bar       & candidate for the Kitaev model        \\
mS24              & 2012 \cite{AK21}                      &             & revised structure                     \\
                  &                                       &             &                                       \\
CrB$_4$           & 2011 \cite{AK17}                      & 1 bar       & distorted 3D boron framework          \\
oP10              & 2012 \cite{AK22}                      &             & misidentified for over 40 years       \\
                  &                                       &             &                                       \\
MnB$_4$           & 2014 \cite{AK28}                      & 1 bar       & distorted 3D boron framework          \\
mP20              & 2014 \cite{Barbara2014,C4CP01339E}    &             & unsolved for over 40 years            \\ \hline \hline
\end{tabular}
\caption{MAISE confirmed predictions with listed ground state
  structures, synthesis pressure, established metastability under
  normal conditions (for phases synthesized at high pressures), key
  features, and general observations.}
\label{Tstr}
\end{table}

CaB$_6$ proved to be the most challenging case in our
structure prediction work. A preliminary ground state search uncovered
several CaB$_6$ polymorphs stabilized by high pressure but none of
them matched the high-pressure XRD patterns obtained in our concurrent
experiments \cite{AK23}. An ES for a larger 28-atom unit cell
eventually converged to a new tI56 structure (SG\#139) with unique
boron building blocks that explained the convoluted XRD data. In
contrast to studies that determined ground states of similar size, the
ES for CaB$_6$ did not use any structural input from experiment, which
makes tI56 one of the largest confirmed crystal structures found truly
'from scratch'. Our follow-up tests for tI56-CaB$_6$, oC88-Li, and
$\gamma$-B$_{28}$ showed that the use of unit cell dimensions
extracted from XRD makes it possible to find the ground state one-two
orders of magnitude faster \cite{AK23}.

Na$_3$Ir$_3$O$_8$ was experimentally observed to transform
into a lower-symmetry phase under pressure. Given the considerable
size of the 56-atom ambient-pressure ground state, we used it to
initialize our ES but did not rely on any high-$P$ experimental data. An
independently obtained mP56 solution (SG\#4) with a dimerized Ir-Ir
network turned out to be in excellent agreement with the collected XRD
patterns \cite{AK36}.

Na$_2$IrO$_3$ structure was originally assigned SG\#15
($C2/c$) \cite{Gegenwart2010}. A simple local optimization revealed
the ground state to have SG\#12 ($C2/m$) in agreement with the
experimental solution established by our colleagues in a joint study
\cite{AK21}. Our RDF analysis helped rationalize the bond
rearrangement resulting in the more stable configuration. The compound
has received considerable attention as a candidate for the realization
of the Kitaev model.

CrB$_4$ \cite{AK17} was first synthesized over 50 years ago and
represented as an oI10 structure (SG\#71). Having determined that
FeB$_4$ is significantly more stable in the related distorted oP10
configuration (SG\#58) \cite{AK16,AK17}, we re-examined CrB$_4$ and
showed oP10 to be the ground state for this compound as well. The
significant distortion of the 3D boron framework was shown to have
little effect on the powder XRD patterns which explained the
mischaracterization of the CrB$_4$ structure. Following electron
diffraction \cite{AK22} and single-crystal XRD \cite{Barbara2013}
measurements confirmed the revised oP10 solution for CrB$_4$.

MnB$_4$ \cite{AK28} was also synthesized over 50 years ago and
tentatively assigned an mS10 (SG\#12) structure. Our ES found a more
stable mP20 (SG\#14) derivative in early 2013. Matching solutions were
obtained independently by several groups around the same
time \cite{Barbara2014,C4CP01339E}.

Our predictive work has shown that crystalline ground states can be
found rather routinely without the need of advanced structure
prediction algorithms if (i) the unit cells have fewer than about 10
atoms; (ii) the search is initialized with related configurations; or
(iii) the search is constrained with unit cell dimensions extracted
from experiment. The ES becomes essential for larger systems,
especially when no prior information is available.

\section{Neural network model construction}
\label{Snn}

In contrast to traditional classical potentials crafted to describe
particular interaction types \cite{Gupta, SCpot,
  Jones463,PhysRevLett.56.632, EAMpot, 0953-8984-5-17-004, AK06,
  ReaxFF,Sinclair1984, Pettifor1999}, common NN models are intentionally
kept devoid of any embedded physics to achieve better transferability
\cite{doi:10.1063/1.4966192}. The NNs' great interpolation power comes
with users' great responsibility to generate proper reference datasets
and perform careful fitting. This section describes key steps for
building general NN interatomic potentials and overviews guidelines
for constructing practical NNs applicable to compound prediction.

\subsection{Reference data generation}

The starting point in NN construction involves choosing a suitable
reference interaction description method and selecting particular
parts of the PES to approximate. Both choices are essential because
NNs inherit the method's systematic/numerical errors and represent the
PES well only in or near the sampled regions. While there are
well-established comparable DFT approximations that can be picked to
describe targeted materials properties
\cite{LDA,PBE,LDA+U1,LDA+U2,vdW1,PhysRevLett.91.126402,SCAN},
automated protocols for generating reference dataset are still being
developed and tested
\cite{PhysRevB.85.045439,AK34,Pt13,AIRSSsampling,Bernstein2019,active-learn-sampling1,active-learn-sampling2,enhanced-sampling}.

As a general principle, it is natural to expose NNs to typical
configurations that will be encountered in intended applications, such
as ground/transition state searches, MD, Monte-Carlo simulations,
vibrational property calculations, etc. In our previous study
dedicated to unconstrained searches \cite{AK34}, we departed from the
popular MD-based scheme and introduced an evolutionary sampling
approach reviewed and generalized further in Section
\ref{Smaisenet}. With the bulk of the diverse dataset created in an
unsupervised fashion, we keep an option open for customized input.

One important recourse discussed in Ref. \cite{AK34} is the
incorporation of equation of state (EOS) data for select structures,
e.g., the dimer, FCC, BCC, HCP, etc., which helps reduce the number of
NN artifacts. We demonstrated \cite{AK34} that inclusion of such
structures with very short and very long interatomic distances has
little effect on the NN description of low-energy structures but
teaches the NN to disfavor unphysical configurations that can be
inadvertently probed in global searches or MD runs. We found this
approach to work better than the common introduction of a repulsive
potential. Another beneficial option is the elimination of structures
that are either too similar to each other or clearly irrelevant. The
reduction of similar structures is performed naturally in our short
evolutionary runs during data generation. The exclusion of structures
with high energy or forces is done during data filtering as detailed
in the next Section. Our typical datasets consist of 86\% of
evolutionary data with 1-8 atoms per unit cell, 12\% of EOS data, and
2\% of structures obtained during evolutionary testing of NN models
(more details in Section \ref{Smaisenet}).

Standard target values taken from DFT calculations are total energies,
atomic forces, and unit cell stresses. In energy training, the outputs
of an atomistic NN model need to be summed up for an entire unit cell
before they can be compared against the corresponding DFT value. In
energy-force training, implemented and examined in our studies in
early 2000's \cite{AK00}, the dataset is expanded dramatically with
more direct information about local environments. Due to the
correlation of forces on nearby atoms according to Newton's
3$^\mathrm{rd}$ law, we randomly pick only 25-50\% of atoms with
non-zero forces in a structure. The resulting ratios of force to
energy data in our studies are at least 7:1.

\subsection{Data filtering and parsing}
\label{SSparsing}

The data processing step allows the user to filter out irrelevant
configurations, earmark structures for training and testing, and parse
atomic environments into NN inputs. These operations can be customized
by choosing flags in the 'setup' file, arranging the data by type into
subdirectories, and specifying Behler-Parrinello (BP) symmetry
functions \cite{PhysRevLett.98.146401} in the 'basis' file.

In data filtering, the ECUT, EMAX, and FMAX flags described in Table
\ref{Ttrainsetup} control the maximum values of energy (enthalpy) and
forces allowed in the database. A single energy cutoff is ill-defined
or not helpful if the database contains entries with different
structure types (clusters or crystal structures), compositions (in
multielement systems), or simulation conditions (pressure
values). Provided that the data is sorted in subdirectories by type,
ECUT and EMAX are applied to the energy per atom within each
subset. These values can be overwritten for a specific subset by
placing a 'tag' file in the corresponding subdirectory. This 'tag'
file can also be used to promote the inclusion of the subset, e.g.,
EOS data, into the training set.

The energy and force cutoff parameters are critical for striking a
balance between the accuracy and the reliability of a NN. It may be
tempting to keep EMAX and FMAX below about 0.5 meV/atom and 1 eV/\AA,
respectively, for exploration of (nearly) stable phases. However, our
tests have shown that such NNs develop numerous artificial minima
easily accessible in MD or structure optimization runs, a problem
known not only for NNs but also for traditional potentials. We have
found that when the cutoff values are raised to 5 eV/atom and 10
eV/\AA, and even higher for EOS data, the NNs lose 1-2 meV/atom in
accuracy but become robust enough to be used in unconstrained
searches.

In data parsing, the idea is to precompute and store NN inputs for
each structure only once to avoid performing this costly operation at
each NN fitting step. The BP symmetry functions used for the
conversion can be easily customized by adjusting the parameters in the
'basis' file. We typically use the set with 51 functions per element
with the cutoff expanded from 6.0 \AA\ to 7.5 \AA\ and the corresponding
$\eta$ parameters rescaled by a factor of 1.25 (as described in our
previous study \cite{AK37}).

The filtering, earmarking, and parsing operations are done in a single
JOBT=30 run. It produces a file for each structure with parsed
energy/force NN inputs and collects statistics on the energy, force,
volume, and RDF distributions in the full dataset.

\subsection{Neural network training}

The default NN implemented in MAISE has a standard feed-forward
architecture with one bias per input or hidden layer. Signals are
processed with hyperbolic activation functions in hidden layers and
with the linear function in the output neuron. Our tests on metallic
alloys have shown comparable performances of one- or two-layer NNs
with the same total number of neurons and insignificant NN accuracy
improvements beyond 20 neurons \cite{AK34}. Based on these
observations, we have adopted the 51-10-10-1 architecture with
$(51+1)\times 10 + (10+1) \times 10 + (10+1)=641$ adjustable
parameters per chemical element.

The filtered and parsed data is split into training and testing sets
with the NTRN and NTST flags, usually at the 9:1 ratio. Data earmarked
for training with 'tag' files in the corresponding subdirectories (see
Section \ref{SSparsing}) has a higher priority to be placed into the
training set.

NN fitting via backpropagation can be performed with BFGS or CG
algorithm as implemented in the GSL. Analytic derivatives of the
weights are used in both energy and force training, with the latter
procedure being slower by a factor of $\sim 3$ per data point. In
order to balance the significance of the energy and force data, the
contributions to the full error function from the mismatches between
the NN and target force component values are multiplied by
0.1 \AA\ (throughout the present work, 'error' represents the
root-mean-square error). The NN weights are initialized randomly or
read in from a previous 'model' file.

\begin{figure}[h]
\centering
\includegraphics[width=0.7\textwidth]{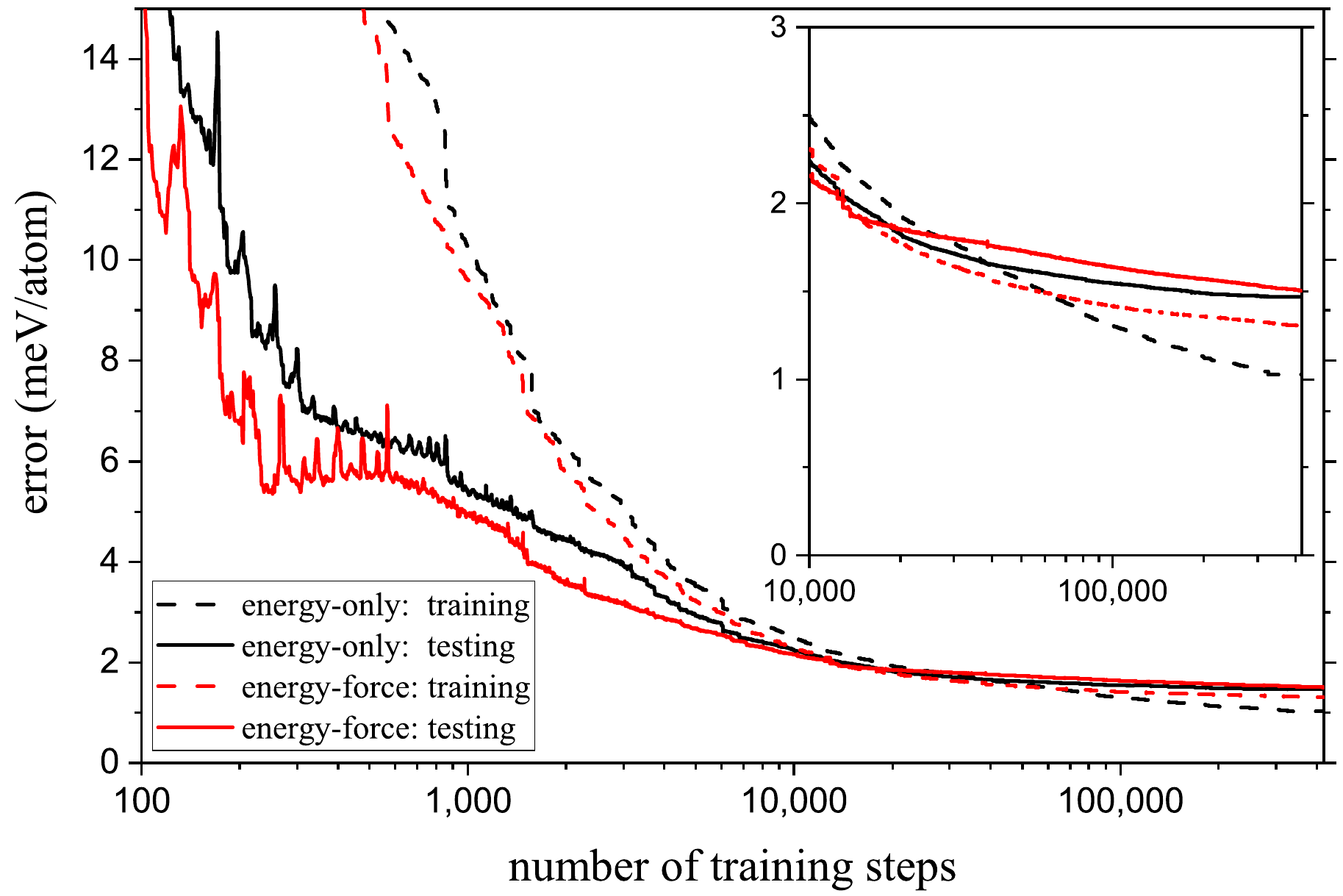}
\caption{Error convergence in optimizations of a Cu-Ag NN model with
  1,880 adjustable parameters for the same binary structure set with
  5,352 energy-only training data (black lines) or 5,352-37,803
  energy-force training data (red lines). The ratio of the training to
  testing data is 9:1. The training errors (dashed lines) are
  higher than the testing errors (solid lines) for the first 20,000
  steps because the training set includes high-energy EOS data. The NN
  trained only on energies displays a sign of overfitting after about
  50,000 steps, while the one trained on energy-force data shows
  comparable training and testing errors (with or without EOS data)
  until the end of the 420,000 optimization run.}
\label{Ptrnconv}
\end{figure}

The optimization is usually carried out for $1-5\times 10^5$
epochs. We have observed that initial weight values have little effect
on the resulting NN accuracy and that NN snapshots saved during an
optimization run provide similar description of EOS, defect energy,
and phonons (Figure S4 in Ref. \cite{AK34}). Overfitting is avoided by
keeping the data to parameter ratio above 10:1 and using $L_2$
regularization with $10^{-8}-10^{-6}$ values. Figure \ref{Ptrnconv}
shows typical rates of convergence in energy and energy-force training
runs.

\subsection{Stratified training}

The construction of NNs for multielement systems in MAISE follows a
stratified scheme introduced and examined in our previous study
\cite{AK34}. It differs from the traditional approach in that we fit
NN weights in a hierarchical fashion from the bottom up, first for
elements, then for binaries, and so on. The intact description of the
subsystems, as the NN is expanded to more elements, is achieved via
the use of a constrained NN architecture. The concept of
stratification has been used in the development of classical and
tight-binding models
\cite{PhysRevB.80.165122,paper02,PhysRevB.84.155119} but has not
received much attention yet in the development of machine learning
potentials.

Under ideal conditions - given a complete basis for representing
atomic environments within a large cutoff sphere, unlimited number of
adjustable parameters and reference data, and a powerful fitting
algorithm - a multielement NN with fully optimized elemental and
interspecies weights is expected to accurately map the PES for all
subsystems. In practice, the use of approximations leads to the
following problem. Suppose one wishes to fit a model describing A, B,
and AB phases given three datasets of A, B, and AB structures. Let's
say that the PES of element A happens to be trivial and can be
approximated with negligible error in the region spanned by the A
data. If one now fits all parameters simultaneously to the full A, B,
and AB dataset the larger error will be distributed across all
elemental and binary systems. In other words, the addition of B and AB
data unphysically alters the description of the elemental A phases. It
should be noted that the constrained NN architecture does account for
the change in the interaction strength between A atoms induced by the
presence of B atoms because the AA/AAA inputs are mixed in with the
AB/AAB/ABB inputs via neurons' non-linear activation functions.

In a study of a particular composition, e.g., MgO, it would not make
much sense to start the parameterization with the elements because
they will not be encountered in charge-neutral forms or relevant
coordinations in MgO structures. With our primary interest in the
exploration of full compositions in multiple binary/ternary metal
alloys, we have relied on the stratified scheme to build sets of
reusable NN models. Our extensive tests have shown that the
constraints in the adopted NN architectures do not introduce any
appreciable errors for the considered chemical systems \cite{AK34}.

In addition to having a more sound foundation, the stratification
procedure significantly accelerates the creation of NN libraries. For
example, the full training of a binary AB model on all A, B, and AB
data takes about the same time as the sequential training of A, B, and
AB models on the corresponding data subsets. However, for an extended
block of A, B, and C elements, the standard approach involves the
fitting of AC and BC NNs from scratch, while the inheritance of A and
B weights in the stratified scheme reduces the total fitting time by
at least a factor of two. The speed-up increases dramatically as more
elements are added and ternary models are built.

Users can choose the full or stratified scheme with the JOBT flag in
the 'setup' file. In the latter case, substituent models should be
placed in the working directory, e.g., 'Cu.dat' and 'Pd.dat' for
fitting the Cu-Pd binary NN, or 'CuPd.dat', 'CuAg.dat', and 'PdAg.dat'
for fitting the Cu-Pd-Ag ternary NN. Presently, MAISE allows for
training NN models with up to three elements. While the treatment of
systems with more elements is possible conceptually, the practical
cost of data generation and parameter optimization becomes expensive.

\subsection{Generalized stratified training}

\begin{figure}[h]
\centering
\includegraphics[width=0.8\textwidth]{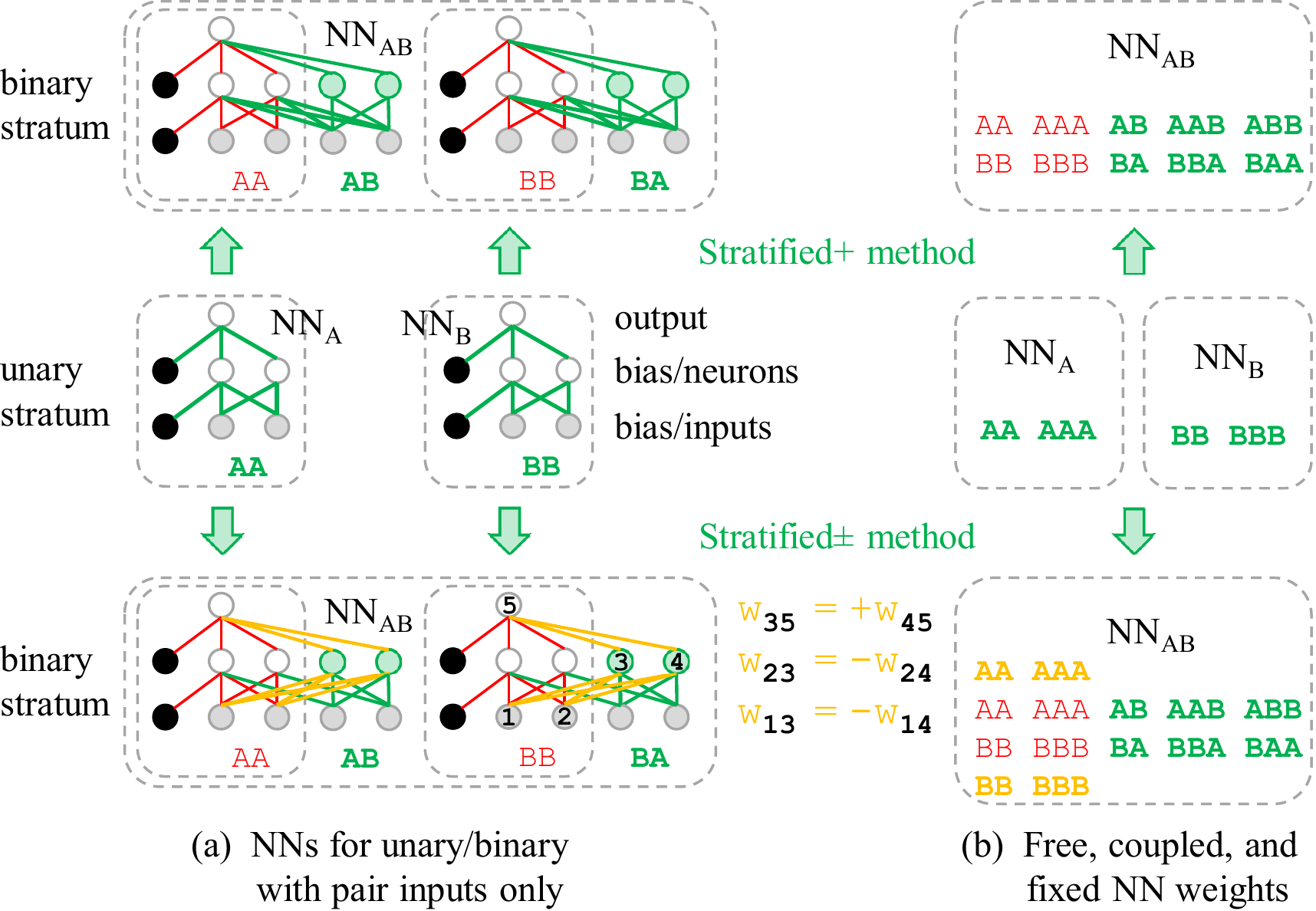}
\caption{Schematic illustration of stratified+ (top row) and
  stratified$\pm$ (bottom row) NN architectures for a binary chemical
  system. The expansion of the original stratified architecture is
  done with the addition of new neurons shown in green. The weights of
  elemental NNs (middle row) are copied and kept fixed in all
  stratified variations. Free, coupled, and fixed weights are shown in
  green, yellow, and red, respectively. (a) Connections in a
  simplified NN with one hidden layer and only pair inputs. The
  partial constraints shown in yellow and explained in the main text
  ensure intact description of the elemental structures. (b)
  Color-coded degrees of weight constraints in NNs with pair and
  triplet inputs. The original and stratified+ schemes have 60\%
  adjustable weights in the first layer in binaries, 11\% in ternaries
  (e.g., only the last one among AA, AB, AC, AAA, AAB, AAC, ABB, ACC,
  ABC, see Ref. \cite{AK34}), and none in quaternaries. The
  stratified$\pm$ architecture can be used for an arbitrary number of
  chemical elements.}
\label{Pstrat}
\end{figure}

In order to extend the stratified procedure to materials with more
complex interactions and an arbitrary number of elements, we have
considered more flexible NN architectures that still preserve the
intact description of the subsystems. Compared to the original
stratified NN layout \cite{AK34}, it involves addition of new neurons,
shown as green units in Figure \ref{Pstrat}, with different connection
patterns and conditions.

The schematic of a 'stratified+' binary NN (top row in
Figure \ref{Pstrat}) illustrates that as long as there are no
connections {\it from} the inputs or neurons in the elemental subnets
{\it to} the inserted neurons, the new adjustable weights do not alter
the signal processing for pure elemental structures. Despite the added
flexibility, the NN still does not allow the proper fitting of
interactions in compounds with more than three chemical
elements. Indeed, the adjustable parts of such NNs involve 60\% of
inputs in binaries (top right box in Figure \ref{Pstrat}), 11\% of
inputs in ternaries (caption of Figure \ref{Pstrat}) and none for
systems with more elements. In our previous discussion \cite{AK34}, we
incorrectly attributed this limitation to the use of pair and triplet
symmetry functions. This restriction is actually imposed not by the
particular geometric representation of the atomic environments
\cite{PhysRevLett.98.146401,PhysRevB.87.184115,2001.11696} but rather
by the NN architecture, and can be lifted as follows.

The 'stratified$\pm$' expansion (bottom row in Figure \ref{Pstrat})
introduces semi-adjustable links even in the inherited parts of the
merged NN. We add neurons in pairs, coupling the two weights incoming
from each subsystem input to have opposite values while coupling the
two outgoing weights to be the same. For a purely elemental structure,
the interspecies input values are zero and the net signal (at neuron 5)
from each elemental input (1) passed through the paired neurons (3\&4)
will be zero as well regardless of the coupled weight {\it
  magnitudes}. For a binary structure, the non-zero binary inputs
multiplied by fully unconstrained weights will unbalance the elemental
signals because of the non-linear nature of the activation function
resulting in a non-zero contribution at neuron 5 that depends on {\it
  both} elemental and binary (semi)adjustable weights.

The set of new partially constrained weights shown in yellow in Figure
\ref{Pstrat} enables the stratified$\pm$ NN to better capture the
screening and charge transfer effects as well as describe interactions
in systems with an unlimited number of species. In a trial
implementation, we imposed the constraint by penalizing the mismatch
between the coupled weights as $\sum_N \sigma (w_{1,N} \pm
w_{2,N})^2$. We have observed no need to adjust the $\sigma$ penalty
factor during the NN optimization, as the differences between coupled
weight magnitudes become negligible after a few dozen training steps;
near the end of optimization, we set the magnitudes to their average
and keep them fixed without any appreciable effect on the error. To
the best of our knowledge, this semi-constrained solution for
systematically expanding NN features has not been considered in the
field of materials modeling. It adds to the collection of alternative
NN architectures proposed in recent years for more general
applications, such as progressive \cite{Rusu2016}, dynamically
expandable \cite{Yoon2018}, and implanted \cite{PhysRevB.97.094106}
NNs.

One way to determine whether the use of the expanded NN architectures
is warranted is to reoptimize the standard stratified NN without any
constraints on the full dataset. A significant reduction in the
training and testing errors would indicate the need for additional NN
flexibility.  In our studies of metal alloys, the error reductions are
usually in the 0-15\% range (e.g., see Figure 4 in
Ref. \cite{AK34}). Our preliminary tests have shown that both
stratified+ and $\pm$ architectures end up with errors about midway
between those in the stratified and full NNs. In order to quantify the
improvements arising from the additional degrees of freedom in each
scheme, we plan to investigate more challenging systems comprised of
different element types in future studies.

\section{MAISE-NET: automated generator of neural networks}
\label{Smaisenet}

Generation of reference structures suitable for tuning machine
learning models has been explored in numerous studies
\cite{PhysRevB.85.045439,PSSB:PSSB201248370,csanyi-C,csanyi-B,Bernstein2019,sample1,sample2,Tribello03042012,jp0757053,QUA:QUA21398,normal-mode-sampling,AIRSSsampling,active-learn-sampling1,active-learn-sampling2,entropy-maximization,enhanced-sampling,Car2019NN}. {\it
  Ab initio} MD has been a particularly popular approach to sample
physically meaningful configurations
\cite{PhysRevB.85.045439,PSSB:PSSB201248370}. In our previous work, we
argued that datasets created with MD might not have the sufficient
representation of diverse environments probed in global structure
searches \cite{AK34}. Our evolutionary sampling protocol proposed in
2017 served as a basis for an unsupervised creation of diverse
datasets, and our NN models trained on such data have been
successfully used in structure prediction \cite{AK37,AK38,AK40}. A
similar approach was developed by Dolgirev {\it et al.}
\cite{Oganov-ES}. Several strategies to improve the mapping of
configuration spaces have been developed in recent years, e.g., normal
mode sampling \cite{normal-mode-sampling}, active learning-based
models \cite{active-learn-sampling1,active-learn-sampling2}, enhanced
sampling \cite{enhanced-sampling}, \textit{ab initio} random structure
searching \cite{AIRSSsampling,Bernstein2019}, and entropy-maximization
approach \cite{entropy-maximization}. A number of studies have shown
the benefit of iterating the generation of data and the
parameterization of models
\cite{PhysRevB.85.045439,AK34,Pt13,AIRSSsampling,Bernstein2019,active-learn-sampling1,active-learn-sampling2,Car2019NN}.

The generalized sampling protocol implemented in the MAISE-NET wrapper
\cite{MAISE-NET} relies on the evolutionary search, structure
analysis, and NN fitting features in MAISE to construct training
datasets in an automated iterative fashion. It has been developed over
several studies to deal with systems of increasing complexity. In our
early investigation of crystal structure phases of relatively simple
Cu-Pd-Ag metals, it was sufficient to generate each of the unary,
binary, and ternary datasets in a single cycle, as the NNs trained on
this data showed robust performance \cite{AK34}. An accurate
description of Cu-Pd-Ag and Au NPs required an iteration to sample
cluster geometries with pretrained NNs \cite{AK38,AK40}. Our ongoing
studies have been dedicated to predicting high-pressure alloy phases
and involve several cycles to include unusual motifs stabilized under
compression.

\begin{figure}[h]
\centering
\includegraphics[width=0.8\textwidth]{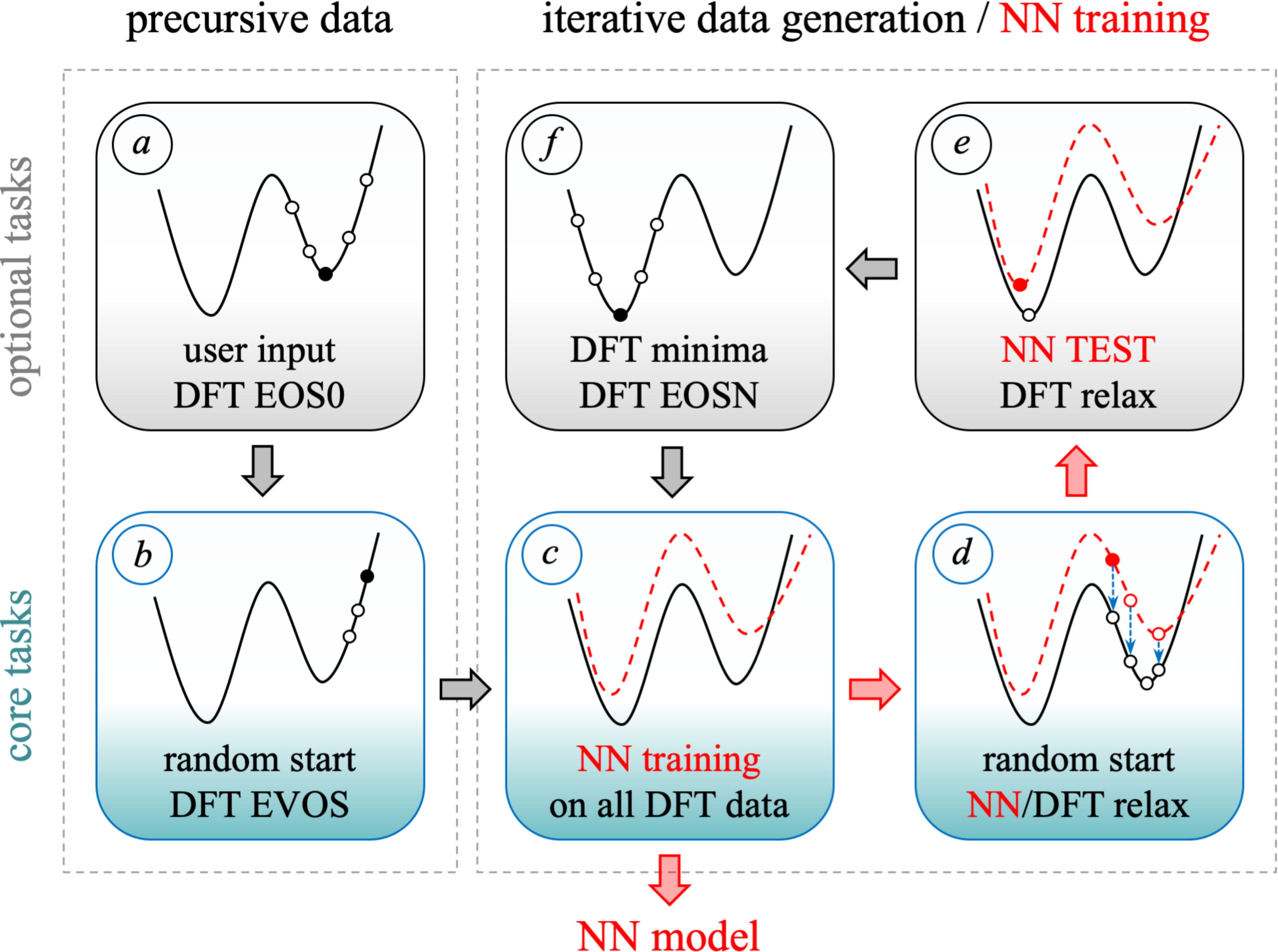}
\caption{A flowchart of the MAISE-NET automated generation of
  reference data and construction of NN models. The core and optional
  tasks are shown in blue and grey boxes, respectively. Black and red
  curves represent the reference DFT PES and its NN approximation,
  respectively. Data produced in steps (a,b) is used to launch an
  iterative process shown in steps (c-f). A detailed description of
  all stages is given in the text.}
\label{Pflowmaisenet}
\end{figure}

An overview of the MAISE-NET operation is presented in Figure
\ref{Pflowmaisenet}. A database construction run starts with building
a precursive dataset followed by cycles of data generation and NN
model training. The complete data generation process is carried out in
multiple steps as follows:

(a) \textit{Basic data generation (optional)}: If instructed by the
user, the script generates a single atom reference and sets of EOS
data for small clusters with 2-4 atoms and select high-symmetry
prototypes preoptimized for the considered element(s). While being
optional, these reference sets, called collectively as EOS0, are
essential for teaching the NN to disfavor configurations with
unphysically short or long interatomic distances.

(b) \textit{Preliminary DFT-level evolutionary sampling}: MAISE-NET
sets up short evolutionary MAISE runs initialized with random
structures. As described in Ref. \cite{AK34}, the local DFT
optimization of each population member for a few ionic steps is
followed by an accurate static evaluation of the energy and forces of
the resulting configuration. The small set of structures in this cycle
0 samples the walls of multiple basins and is sufficient for a rough
approximation of the PES.

(c) \textit{NN model training}: The collection of all available
high-accuracy DFT data is parsed and a NN model is built. Various
system- and cycle-dependent fitting specifications can be defined in
the setup file, e.g., the energy or energy-force training type, the
number of steps for each training type, etc.

(d) \textit{NN-driven generation of DFT data}: MAISE-NET launches
MAISE evolutionary runs to randomly generate and locally optimize new
structures using the latest NN model. Compared to step (b), it proved
to be unnecessary to proceed beyond the first ES generation because
small unit cells with 1-8 atoms have a chance to converge to the
global minimum with full local optimization affordable at the NN
level. After the uniqueness of the obtained minima is verified through
the structure comparison feature in MAISE and they are accepted in the
pool based on a weighting factor favoring low-enthalpy structures, the
corresponding relaxation paths are examined to extract several
intermediate structures per minimum. The target total number of
generated structures per cycle, referred to as 'EVOS' data, is
specified in the setup file.

(e) \textit{NN model test and TEST data (optional)}: If instructed by
the user, MAISE-NET launches a proper evolutionary ground state search
using the NN model trained in step (c). The resulting NN-based minima
are then optimized at the DFT level. A detailed report is compiled on
the symmetry and enthalpy of the resulting minima at the NN and DFT
levels. The user has an option to include the DFT energies and forces
of the NN- and DFT-based minima into the dataset for the NN training
in subsequent cycles. The data will be added to the collection of
training data as 'TEST' data. Although generation of the TEST data
during the model construction run is optional, the script has the
feature to perform this evolutionary search test for an existing NN
model as an independent functionality.

(f) \textit{DFT EOSN data generation (optional)}: If instructed by the
user, a small set of EOS data is generated for unique DFT-optimized
minima obtained in each cycle and added to the pool of training data
for the next cycles as 'EOSN' data.

\begin{figure}[h]
\centering
\includegraphics[width=\textwidth]{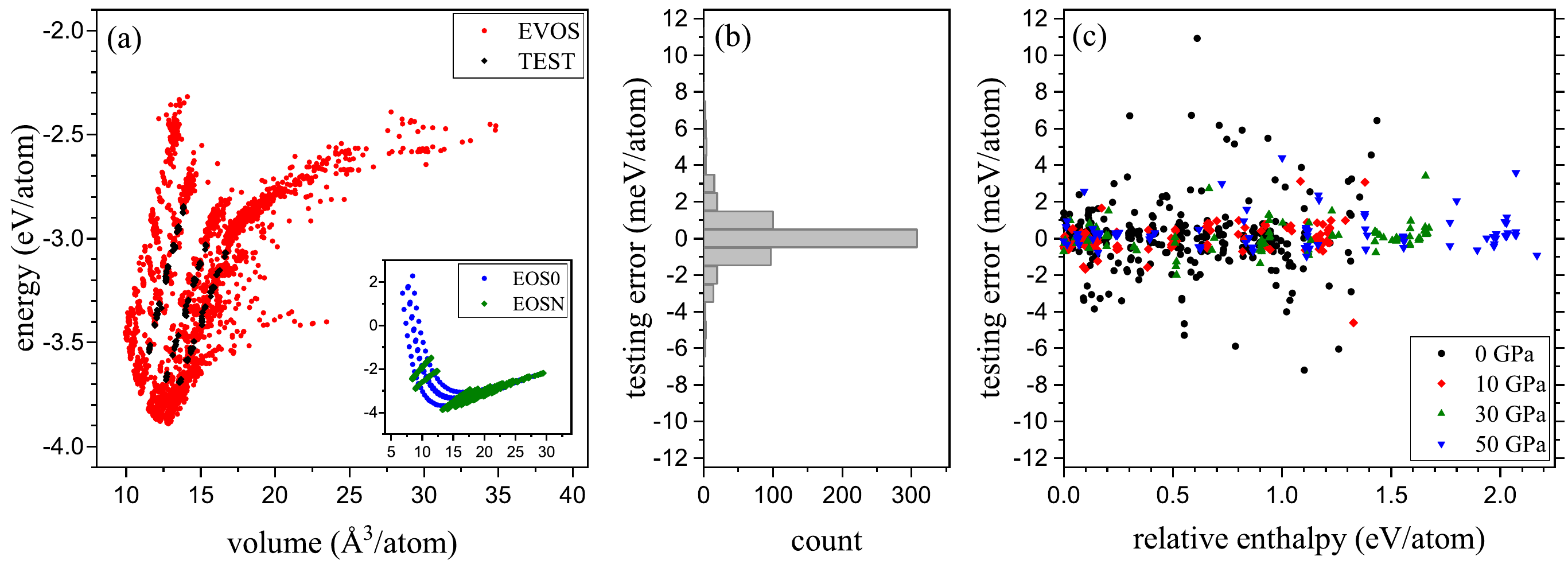}
\caption{Characteristics of the generated dataset and accuracy of the
  constructed NN model for the Cu-Ag binary. (a) Distribution of the
  generated dataset for various data types: EVOS (evolutionary data),
  TEST (NN model testing results), EOS0 (basic data generated before
  zeroth cycle), and EOSN (EOS data generated during the evolutionary
  run). (b) Histogram of energy errors the testing set. (c)
  Distribution of testing errors as a function of the enthalpy
  difference relative to the lowest-enthalpy phase at each pressure.}
\label{Pmaisenet}
\end{figure}

Steps (c) through (f) are repeated for a user-specified number of
cycles, with a NN model trained from scratch on all collected DFT data
at the end of each iteration. The run can be terminated or extended by
the user at each iteration depending on whether a satisfactory
accuracy for the NN model is achieved. While steps (a), (e), and (f)
are optional, our tests for elemental, binary, and ternary metal
systems have indicated that addition of these datasets significantly
improves the NN model suitability for ground state searches in terms
of accuracy and reliability. Generation of a typical training dataset
of $\sim5,000$ structures with the MAISE-NET script required roughly 20,
30, and 40 thousand CPU hours of DFT calculations for elemental,
binary, and ternary metallic compounds, respectively. The higher DFT
calculation cost for each binary and ternary systems is primarily due
to the increased number of DFT calculations for structures with larger
unit cells. Figure \ref{Pmaisenet} illustrates the distribution of data
and NN accuracy for Cu-Ag.

The end-to-end NN construction depends on a large number of parameters
and can be a daunting task for new users. With this in mind, we have
developed MAISE-NET to have the following features.

{\it Easy customization} MAISE-NET can be run with both Python 2 and 3
version families out-of-the-box without requiring any external modules
to be installed. It includes well-tested 'setup' templates for
developing elemental, binary, and ternary NN models. All key
functionalities can be tuned by adjusting 'setup' parameters listed in
Table \ref{Tmaisenetsetup}. Upon detection of user-provided NN models
for the relevant subsystems, the script performs the NN fitting in the
stratified fashion.

{\it Complete automation} Once the run is configured, the data
generation and NN construction can proceed without any further user
input or supervision. In particular, the script makes sure that DFT
data is collected only from successfully finished VASP calculations.

{\it Full transparency} An extensive set of messages is produced and
sent into standard output/screen to notify the user about the progress
of the run. The most important messages are saved in the 'output.dat'
file. A comprehensive summary is generated to give the user a detailed
account about the generated dataset.

\section{Library of neural network models}
\label{Slib}

A library of select NN and empirical potentials is provided with the
distribution in the 'models/' directory. Model file names specify the
interaction type (a NN or a traditional potential), dimensionality of
the data used to parameterize the model (0 for crystals and clusters
or 3 for crystals only), and the generation/version number. Model file
headers list information about models' authorship, architecture,
performance, etc. The body of the NN files contain bias and weight
values. Finally, the end of the files specifies the symmetry function
basis chosen for the model.

\begin{figure}[h]
\centering
\includegraphics[width=0.7\textwidth]{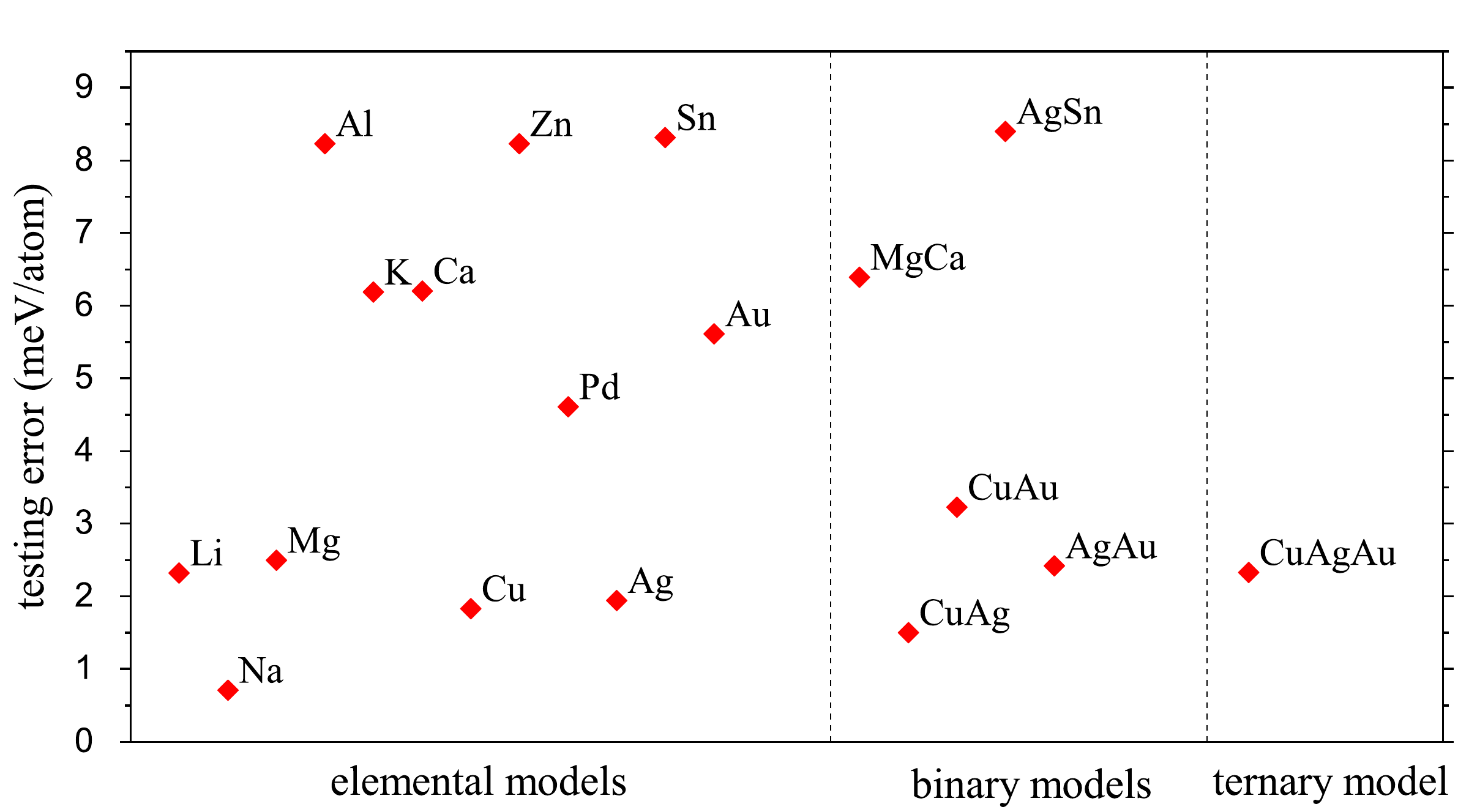}
\caption{Testing error for available second-generation NN interatomic
  potentials constructed with the MAISE-NET script.}
\label{Pliberr}
\end{figure}

We have recently started building a new generation of NN models (gen2)
for a large set of metals to allow the prediction of stable alloys
under ambient and high pressures. The use of MAISE-NET with standardized
settings ensures that we can create a library of models in the
stratified fashion. Figure \ref{Pliberr} shows the accuracy of the new
generation of NNs tested to perform well in structure searches up to
30 GPa pressure.

\section{Neural network benchmarks and predictions}
\label{Snntest}

\subsection{Efficiency of NN calculations}

\begin{figure}[h]
\centering
\includegraphics[width=\textwidth]{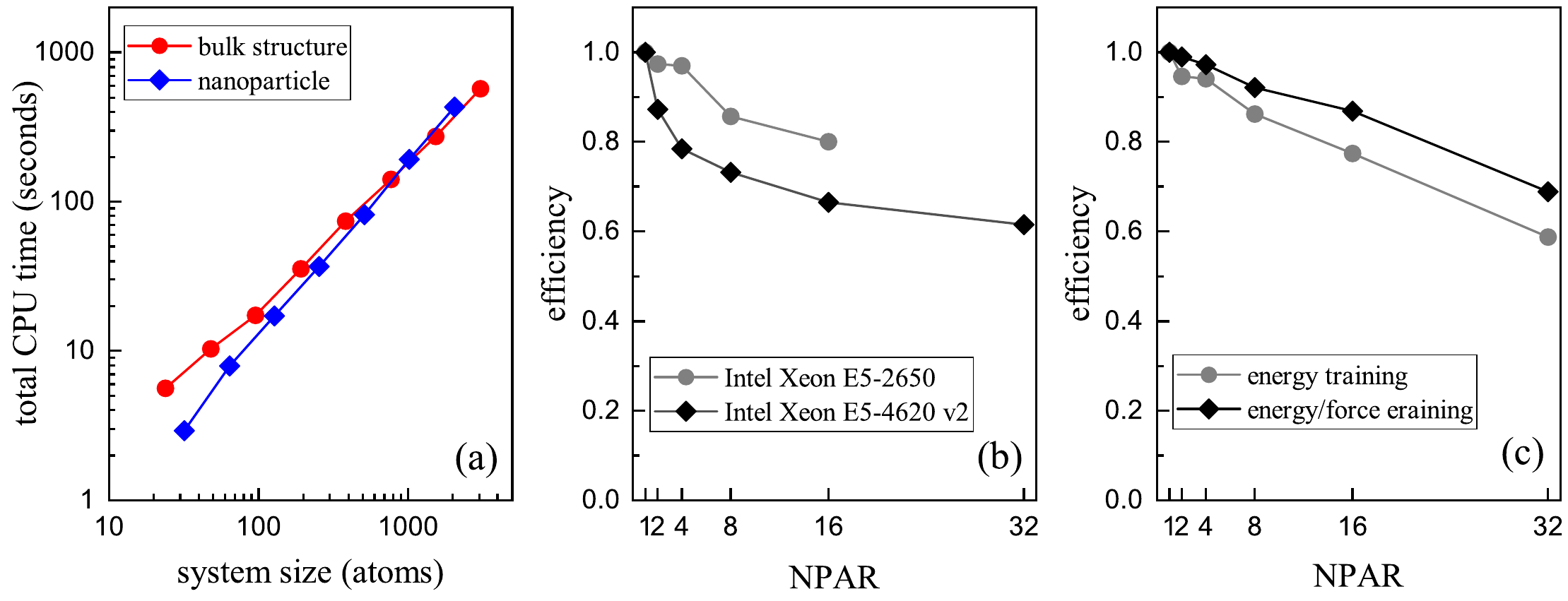}
\caption{MAISE performance in structure relaxation and NN model
  training. (a) Total CPU time for the relaxation of bulk (red
  circles) and NP (blue diamonds) Au structures performed on a 32-core
  Intel Xeon Gold 5218 @ 2.30 GHz compute node. (b) Parallelization
  efficiency of MAISE for local optimization of a 1024-atom Au crystal
  structure computed on a 16-core Intel Xeon E5-2650 @ 2.00 GHz (grey
  circles) and a 32-core Intel Xeon E5-4620 v2 @ 2.60 GHz (black
  diamonds). The dynamic allocation in OpenMP helps distribute the
  load for processing atoms with different numbers of neighbors. (c)
  Parallelization efficiency of training a NN model on energy-only
  (grey circles) and energy-force (black diamonds) data, performed on a
  32-core Intel Xeon Gold 5218 @ 2.30 GHz compute node.}
\label{Peffic}
\end{figure}

Benchmarking results reported in our previous studies
\cite{AK34,AK37,AK38,AK40} have demonstrated the levels of speed and
accuracy generally expected from the constructed NN models. For
systems with 50-100 atoms, calculations performed with the order-$N$
NNs were found to be $10^4-10^5$ times faster than with the
order-$N^3$ DFT and about $10^2$ times slower than the order-$N$
empirical potentials \cite{force0,force1,force2,force3}. The two most
demanding computational tasks, the NN training and the NN use in
structure simulations, are parallelized with OpenMP over the total
number of structures in the reference dataset in the former case and
over the number of atoms in the latter one. Figure \ref{Peffic}
illustrates that the parallelization efficiency is system-dependent
and can be up 90\% on 16 cores and 70\% on 32 cores.

\subsection{Accuracy of NN models}

\begin{figure}[h]
\centering
\includegraphics[width=0.8\textwidth]{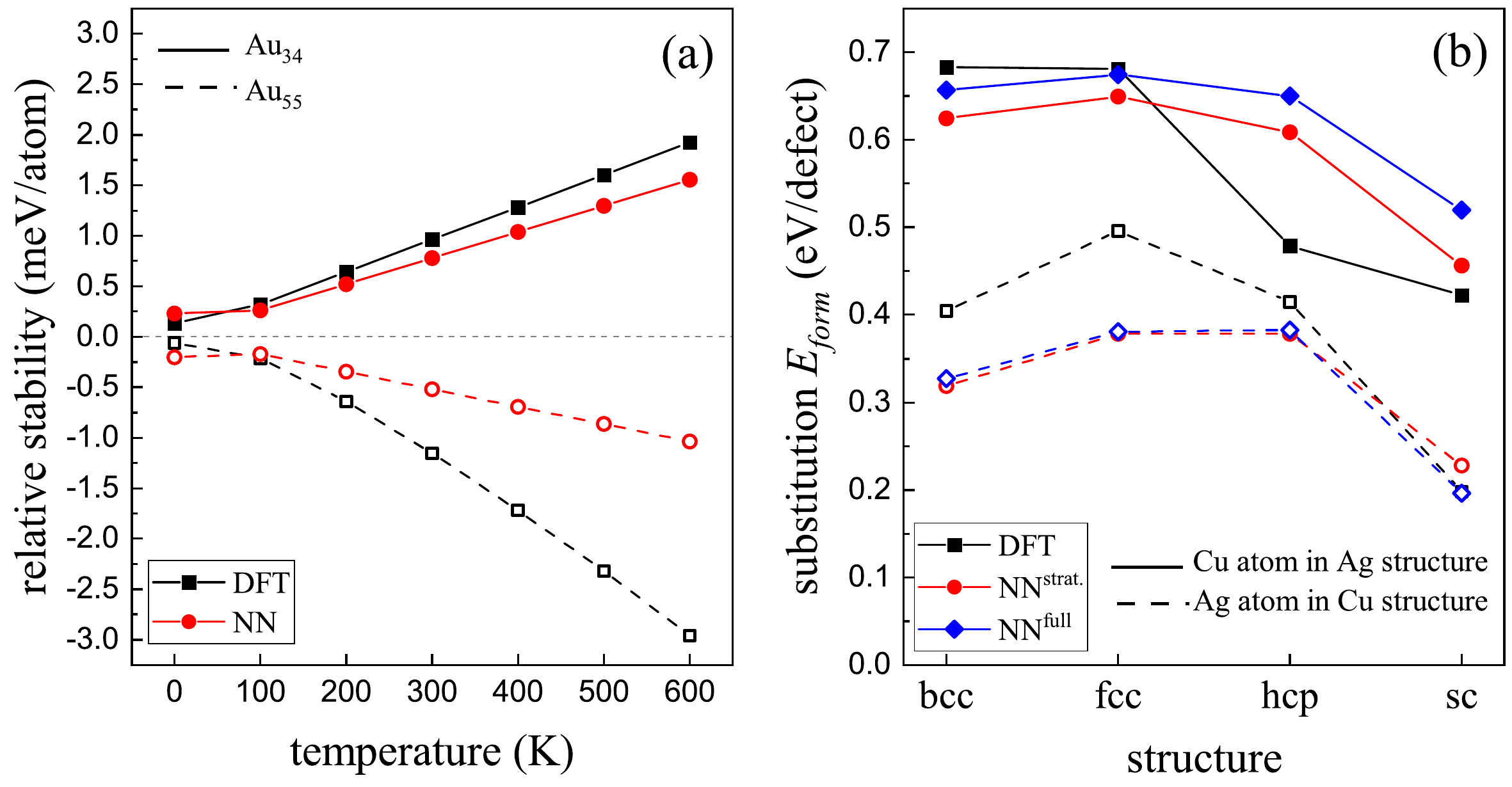}
\caption{Accuracy of the NN models constructed with MAISE in the
  evaluation of different properties. (a) Vibrational entropy
  corrections to the relative stability between two low-energy NPs for
  two sizes, Au$_{34}$ and Au$_{55}$\cite{AK40}. (b) Substitutional
  defect formation energies for Cu-Ag binary structures evaluated with
  stratified (red cirles) and full (blue diamonds) NN models and
  compared to the DFT values (black squares).}
\label{Pnnper}
\end{figure}

As overviewed in Section \ref{Slib}, the overall accuracy for most
developed models ranges between 2 and 10 meV/atom in the considered
systems with up to three metals. The DFT formation defect energies are
typically reproduced within 0.1-0.2 eV/defect (see Figure \ref{Pnnper}),
which is consistent with the NN errors per atom (see discussion in
Ref. \cite{AK34}). The accurate description of forces with the NNs
allows one to identify dynamically unstable structures and obtain
accurate evaluations of relative phase stability at elevated
temperatures by including vibrational entropy corrections
(Figure 5 in Ref. \cite{AK40} and Figure \ref{Pnnper} in the present
work). It has been encouraging to observe practically the same
accuracy of NNs trained in the full and stratified fashions.

\subsection{Comparison of NNs and empirical potentials}

The most important quality measure of NNs developed in our studies
has been their performance in structure prediction
\cite{AK37,AK38,AK40}. We carried out a systematic comparison of NN
models and traditional potentials against the DFT, which appears to be
the largest of the kind so far, to quantify their ability to resolve
low-energy configurations in global structure searches \cite{AK40}. By
examining up to 60 lowest-energy candidates for each size in the
Au$_{30}$-Au$_{80}$ NP range, we established that NNs (6.5 meV/atom
accuracy) are far better suited to guide {\it ab initio} ground state
search than the Gupta, Sutton-Chen, or embedded atom model (estimated
30 meV/atom accuracy). The large number of NP configurations with
close energies makes it difficult to conclusively determine the DFT
minimum with either approach but the use of the NNs instead of the
traditional potentials reduces the number of structures needed to be
re-evaluated at the DFT level by at least 1-2 orders of magnitude.
Moreover, the good correspondence between the NN and DFT atomic forces
allowed us to introduce a hybrid NN+DFT approach that significantly
improves the search reliability. Application of NN models with 2-10
meV/atom accuracy to bulk crystals is expected to be far more
effective for identifying the DFT ground states because of the simpler
PES near the global minimum in systems without surfaces.

\subsection{NN-based structure predictions}

First practical applications of NNs in structure prediction with MAISE
have led to identification of more stable Au NP configurations
\cite{AK40} and new synthesizable Mg-Ca bulk phases \cite{AK37}. In
both studies, NNs were used to describe interatomic interactions
during unconstrained searches and then select candidate structures
were evaluated with DFT approximations.

Au NPs have been subject of numerous studies because of their
appealing catalytic properties
\cite{Astruc2004,Ricci2006,Janssens2007,Turner2008}. A variety of
structure optimization and interaction description methods have been
used to determine ground state NP configurations with up to 300 atoms
as reviewed in our study \cite{AK40}. Small Au clusters have
been shown to adopt unusual stable morphologies, such as the Au$_{32}$
hollow fullerene or the Au$_{34}$ pyramid
\cite{Johansson2004,npann1}. We performed NN-based ESs for
Au$_{30}$-Au$_{80}$ and found our best configurations at the DFT level
to have either matching or better energy compared to all previously
reported Au NPs. In particular, we identified more stable
configurations for sizes 34, 38, and 55. Considering the amount of
work dedicated to magic-size clusters with 55 atoms, it was surprising
to uncover a new Au$_{55}$ configuration 3.6 meV/atom lower in energy
than all putative ground states described in the literature
\cite{AK40}.

Mg alloys have been widely explored because of their potential
applications as structural materials in automotive and aircraft
industries \cite{Mgapp}. The Mg-Ca binary system has only one known
compound, Mg$_2$Ca, with the C14 Laves structure. In our joint study
\cite{AK37}, the full Mg-Ca composition was scanned with the
PyChemia's minima hopping search engine \cite{FireflyPyChemia} using
MAISE as an external NN module. At the ambient pressure, we identified
several phases close to stability at zero temperature: C15/C36 Laves
structures at the 2:1 composition and oS36/mS18 at the 7:2
composition. We demonstrated that the vibrational entropy contribution
makes these phases more stable and they could be overlooked materials
synthesizable at high temperature. At high pressures, B2-MgCa and
cF16-Mg$_3$Ca become thermodynamically stable and are expected to form
below 10 GPa. We are not aware of any earlier reports of new
synthesizable compounds predicted with global structure searches based
on NN interatomic potentials.

\section{Summary}

In this work, we have reviewed notable predictions and present
capabilities of MAISE. The list of eight crystal structure predictions
made at the DFT level and confirmed in concurrent or following
experiments is presented in Section \ref{Sstr}. The identification of
possible synthesizable Mg-Ca phases with global structure searches at
the NN level\cite{AK34}, which appears to be the first example of new
thermodynamically stable crystalline compounds predicted in this
fashion, is discussed in Section \ref{Sstr}. Key aspects of the
evolutionary optimization implemented in MAISE for crystals, films, and
clusters are described in Section \ref{Sevo}.

The main feature of the package is the construction of NN interaction
models for use in structure prediction (Section \ref{Snn}). We outline
our protocols for configuration space sampling and NN training that
ensure the robustness of the DFT PES mapping. In particular, we
introduce expanded stratified schemes that allow the construction of
NN models in a hierarchical fashion for an arbitrary number of
chemical elements. All stages of the iterative NN development are
handled with an automated MAISE-NET wrapper (Section \ref{Smaisenet}). The
script has been used in our ongoing effort to build a new generation
of NN models (Section \ref{Slib}). So far, NNs for 12 metals, 5 binary
alloys, and 1 ternary alloy with an accuracy in the 2-9 meV/atom
range have been tested in unconstrained structure searches at
pressures up to 30 GPa. Section \ref{Ssim} illustrates MAISE and NN
performance in local structure relaxations, MD simulations, and phonon
calculations. MAISE, MAISE-NET, and developed NN models are available for
download on Github \cite{MAISE,MAISE-NET}.

\section{Acknowledgment}
We acknowledge the NSF support (Award No. DMR-1821815) and the Extreme
Science and Engineering Discovery Environment computational resources
\cite{XSEDE} (NSF award No. ACI-1548562, project No. TG-PHY190024). We
thank Igor Mazin, G\'{a}bor Cs\'{a}nyi, and Michele Ceriotti for
insightful discussions.

\newpage
\begin{appendices}
\renewcommand{\thetable}{\Alph{section}\arabic{table}}

\section{Setup parameters for various MAISE features}
\label{Sapp}

This section lists key setup parameters for evolutionary global
structure optimization, local structure optimization, MD simulations,
data parsing, and NN training with MAISE as well as key setup
parameters for the automated NN model construction with MAISE-NET. The
following tables include a minimal set of parameters, i.e., those
which need to have a defined value for the code to operate properly.

\begin{table}[H]
\centering
\begin{tabular}{p{0.1\columnwidth}|p{0.8\columnwidth}}\hline\hline
flag & description            \\ \hline
JOBT & job type: structure relaxation (20) \\
NPAR & number of cores for parallel run \\
NDIM & dimensionality of the unit cell: crystal (3); cluster (0) \\
MITR & maximum number of cell optimization steps \\
RLXT & cell optimization type: force only (2); full cell (3); volume (7)\\
PGPA & external pressure in GPa\\
ETOL &  total energy difference  for cell optimization convergence \\ 
COUT & output options E: final (00); first/final (01); all steps (02);    \\
     & EF: final (10); first/final (11); all steps (12) \\
MINT & minimizer type: BFGS2 (0); CG-FR (1); CG-PR (2); steepest descent (3) \\ \hline\hline
\end{tabular}
\caption{Setup parameters for local structure optimization.}
\label{Trelaxsetup}
\end{table}

\begin{table}[H]
\centering
\begin{tabular}{p{0.1\columnwidth}|p{0.8\columnwidth}}\hline \hline
flag  & description                                                       \\ \hline
JOBT  & job type: molecular dynamics (21)                                 \\
MDTP  & MD run type: $NVE$ (10); $NVT$ (20); $NPT$ (30); isobaric-isothermal (40) \\
NPAR  & number of cores for parallel run                                 \\
TMIN  & starting temperature of the simulation                           \\
TMAX  & final temperature of the simulation                             \\
TSTP  & temperature increment during the simulation                      \\
DELT  & integration time step in fs                                       \\
NSTP  & number of integration steps per temperature                      \\
 CPLT   & thermostat coupling constant                                     \\
CPLP  & barostat coupling constant                                        \\
ICMP  & isothermal compressibility in 1/GPa                               \\ \hline \hline
\end{tabular}
\caption{Setup parameters for MD simulations.}
\label{Tmd}
\end{table}

\begin{table}[H]
\centering
\begin{tabular}{p{0.1\columnwidth}|p{0.8\columnwidth}}\hline\hline
flag & description                                                       \\ \hline
JOBT & job type: phonon calculations (22)                               \\
DISP & size of the displacement made to each atom in \AA\           \\
NPAR & number of cores for parallel run                                   \\
NDIM & dimensionality of the unit cell: crystal (3); cluster (0) \\ \hline\hline
\end{tabular}
\caption{Setup parameters for phonon calculations.}
\label{Tphonsetup}
\end{table}

\begin{table}[H]
\centering
\begin{tabular}{p{0.1\columnwidth}|p{0.8\columnwidth}}\hline\hline
flag & description                                                       \\ \hline
JOBT  & evolutionary search: run (10); soft exit (11); hard exit (12); analysis (13) \\
NMAX  & maximum number of atoms \\
MMAX  & maximum number of neighbors within cutoff radius \\
NSPC  & number of species types \\
TSPC  & species types \\
ASPC  & atom number of each species in evolutionary searches \\
CODE  & MAISE-INT (0); VASP-EXT (1); MAISE-EXT (2) \\
QUET  & queue type: torque (0); slurm (1)\\
NDIM  & structure type: crystal (3); film (2); cluster (0) \\
LBOX  & box size for cluster calculations (ignored for crystals) \\
NPOP  & population size \\
SITR  & starting iteration \\
NITR  & number of iterations \\
TINI  & starting options if SITR=0 \\
TIME  & max time per relaxation \\
PGPA  & pressure in GPa \\
DENE  & energy/atom window for selecting distinct structures \\
SCUT  & RDF difference for selecting distinct structures  \\
TETR  & random using TETRIS \\
PLNT  & seeded \\
PACK  & biased \\
BLOB  & random using blob shape \\
MATE  & crossover using two halves \\
SWAP  & crossover using core-shell \\
RUBE  & Rubik's cube operation \\
REFL  & symmetrization via reflection \\
INVS  & symmetrization via inversion \\
CHOP  & chop to make facets \\
MUTE  & distortion \\
ELPS  & cluster ellipticity \\
MCRS  & crossover:  mutation rate  \\
SCRS  & crossover:  swapping rate \\
LCRS  & crossover:  mutation strength for lattice vectors \\
ACRS  & crossover:  mutation strength for atomic positions \\
SDST  & distortion: swapping rate \\
LDST  & distortion: mutation strength for lattice vectors \\
ADST  & distortion: mutation strength for atomic positions \\
SEED  & starting seed for the random number generator (0 for system time) \\ \hline\hline
\end{tabular}
\caption{Setup parameters for evolutionary search.}
\label{Tevosetup}
\end{table}

\begin{table}[H]
\centering
\begin{tabular}{p{0.1\columnwidth}|p{0.8\columnwidth}} \hline\hline
 flag  &  description \\ \hline
 JOBT  &  job type: data parsing (30) \\
 NPAR  &  number of cores for parallel training or cell simulation  \\
 TEFS  &  parsing for: E (0); EF (1) \\
 FMRK  &  fraction of atoms that will be parsed to use for EF training \\
 NSPC  &  number of element types for dataset parsing and training \\
 TSPC  &  atomic number of the elements specified with NSPC tag \\
 NSYM  &  number of the BP symmetry functions for parsing data \\
 NCMP  &  the length of the input vector of the neural network \\
 ECUT  &  parse only this fraction of lowest-energy structures (from 0 to 1) \\
 EMAX  &  maximum energy from the lowest-energy structure that is parsed \\
 FMAX  &  will not parse data with forces larger than this value \\
 RAND  &  random seed for the parsing: time (0); seed value (+); no randomization (-) \\
 DEPO  &  path to the DFT datasets to be parsed \\
 DATA  &  location of the parsed data to write the parsed data \\ \hline \hline
\end{tabular}
\caption{Setup parameters for data parsing.}
\label{Tparsesetup}
\end{table}

\begin{table}[H]
\centering
\begin{tabular}{p{0.1\columnwidth}|p{0.8\columnwidth}} \hline\hline
 flag  &  description \\ \hline
 JOBT  &  training type: full training (40); stratified training (41) \\
 NPAR  &  number of cores for parallel training  \\
 MINT  &  the optimizer algorithm for neural network training \\
 MITR  &  number of the optimization steps for training \\
 ETOL  &  error tolerance for training \\
 TEFS  &  training target value: E (0); EF (1) \\
 NSPC  &  number of element types for dataset parsing and training \\
 TSPC  &  atomic number of the elements specified with NSPC tag \\
 NSYM  &  number of the BP symmetry functions for parsing data \\
 NCMP  &  the length of the input vector of the neural network \\
 NTRN  &  number of structures used for training (negative number means percentage) \\
 NTST  &  number of structures used for testing (negative number means percentage) \\
 NNNN  &  number of hidden layers (does not include input vector and output neuron) \\
 NNNU  &  number of neurons in hidden layers \\
 NNGT  &  activation function type for the hidden layers' neurons: linear (0); tanh (1) \\
 LREG  &  regularization parameter \\
 SEED  &  rand seed for generating NN weights (0 for system time) \\
 DATA  &  location of the parsed data to read from for training  \\
 OTPT  &  directory for storing model parameters in the training process \\
 EVAL  &  directory for model testing data \\ \hline\hline
\end{tabular}
\caption{Setup parameters for training of NN models.}
\label{Ttrainsetup}
\end{table}

\begin{table}[H]
\centering
\begin{tabular}{p{0.1\columnwidth}|p{0.8\columnwidth}}\hline\hline
 flag & description \\ \hline
 JOBT & job type: basic data generation (80); evolutionary data generation (81); \\
      & test run (87); pause (88); exit (89) \\
 TSPC & atomic number of the elements \\
 QUET & queue type: torque (0); slurm (1); IBM-lsf (2) \\
 LBOX & unit cell size: should be non-zero for BASIC data \\
 MAXJ & maximum number of DFT jobs to be submitted at once \\
 ECUT & energy cut-off for DFT (0 = VASP default) \\
 PREC & precision of the DFT run (e.g., norm,acc) \\
 KDNS & K-mesh density for DFT runs \\
 SMER & VASP ISMEAR \\
 SIGM & VASP SIGMA (for REFS and CLST data will be set to 0.01) \\
 LREG & regularization parameter\\
 NNNU & number of neurons in hidden layers \\
 NNGT & activation function type for the hidden layers' neurons: linear (0); tanh (1) \\
 NPAR & number of cores for parallel parsing \\
 NSYM & number of the BP symmetry functions for parsing data \\
 RCUT & BP symmetry function cut-off radius: 6 \AA\ (0); 7.5 \AA\ (1)  \\
 FMRK & ratio of atomic forces for training \\
 SITR & starting cycle (0 for full run) \\
 NITR & final cycle \\
 DATA & desired number of structures per cycle (1+) \\
 aspc & list of number of atoms/unit cell (for cycle 0) \\
 npop & list of population size for evolution runs (for cycle 0) \\
 mitr & number of training steps (for cycle 0) \\
 tefs & type of training at each round \\
 ASPC & list of number of atoms/unit cell (for cycles 1+) \\
 NPOP & population size for evolution runs (for cycles 1+) \\
 ITER & relaxation steps for NN-based search \\
 MITR & number of training steps (for cycles 1+) \\
 TEFS & type of training at each round (for cycles 1+) \\
 EXTR & extended force training factor when cycle = NITR \\ \hline \hline
\end{tabular}
\caption{Setup parameters for automated model construction with MAISE-NET.}
\label{Tmaisenetsetup}
\end{table}

\end{appendices}

\newpage



\bibliographystyle{unsrt}
\bibliography{ref.bib}



\end{document}